\documentclass[journal]{new-aiaa}
\usepackage[utf8]{inputenc}
\usepackage{textcomp}

\usepackage{graphicx}
\usepackage{amsmath}
\usepackage[version=4]{mhchem}
\usepackage{siunitx}
\usepackage{longtable,tabularx}
\usepackage{amsmath, bm}
\usepackage{thmtools} 

\usepackage{subcaption}
\usepackage{gensymb}
\usepackage{algorithm}
\usepackage[noend]{algpseudocode}
\usepackage{color,soul}         

\usepackage{fancyhdr}
\usepackage{multicol}

\usepackage{adjustbox}
\usepackage{booktabs}

\usepackage{tikz}
\usetikzlibrary{external}
\usepackage{glossaries} 
\makeglossaries

\hypersetup{pdfborder={0 0 0}}
\usepackage{makecell}

\title{Comparing Behavioural Cloning and Reinforcement Learning for Spacecraft Guidance and Control Networks}

\author{Harry Holt\footnote{Research Fellow, Advanced Concepts Team, European Space Research and Technology Centre (ESTEC)}, 
Sebastien Origer\footnote{Young Graduate Trainee, Advanced Concepts Team, European Space Research and Technology Centre (ESTEC)}
and Dario Izzo\footnote{Scientific Coordinator, Advanced Concepts Team, European Space Research and Technology Centre (ESTEC)}}
\affil{Advanced Concepts Team, European Space Research and Technology Centre (ESTEC),\\ ESA, Noordwijk, The Netherlands}

\begin{document}

\maketitle


\newacronym{OCP}{OCP}{optimal control problem}
\newacronym{TPBVP}{TPBVP}{two-point boundary value problem}
\newacronym{MPBVP}{MPBVP}{multi-point boundary value problem}
\newacronym{COV}{COV}{calculus of variations}
\newacronym{PMP}{PMP}{Pontryagin's minimum principle}

\newacronym{OI}{OI}{orbit insertion}
\newacronym{IC}{IC}{initial conditions}
\newacronym{OD}{OD}{orbit determination}
\newacronym{EX}{EX}{thrust execution}

\newacronym{gcnet}{G\&CNET}{guidance \& control network}
\newacronym{BC}{BC}{behavioural cloning}
\newacronym{RL}{RL}{reinforcement learning}
\newacronym{IL}{IL}{imitation learning}
\newacronym{uav}{UAV}{unmanned air vehicles}

\newacronym{NN}{NN}{neural network}
\newacronym{ANN}{ANN}{artificial neural network}

\newacronym{MDP}{MDP}{Markov decision process}
\newacronym{AC}{AC}{actor-critic}
\newacronym{PPO}{PPO}{proximal policy optimisation}
\newacronym{TRPO}{TRPO}{trust-region policy optimisation}
\newacronym{ReLU}{ReLU}{rectified linear unit}
 
\newacronym{GNC}{GNC}{guidance, navigation and control}
\newacronym{GC}{G\&C}{guidance and control}

\newacronym{MC}{MC}{Monte Carlo}
\newacronym{ZOH}{ZOH}{zero-order hold}

\newacronym{SDF}{SDF}{signed distance function}
\newacronym{SOI}{SOI}{sphere of influence}

\newacronym{MPC}{MPC}{model predictive control}
\newacronym{SIREN}{SIREN}{sinusoidal representation network}

\newacronym{bgoe}{BGOE}{backward generation of optimal examples}

\glsresetall

\begin{abstract}

\Glspl{gcnet} provide a promising alternative to on-board \gls{GC} architectures for spacecraft, offering a differentiable, end-to-end representation of the guidance and control architecture. When training \glspl{gcnet}, two predominant paradigms emerge: \gls{BC}, which mimics optimal trajectories, and \gls{RL}, which learns optimal behaviour through trials and errors. Although both approaches have been adopted in \gls{gcnet}-related literature, direct comparisons are notably absent. To address this, we conduct a systematic evaluation of \gls{BC} and \gls{RL} specifically for training \glspl{gcnet} on continuous-thrust spacecraft trajectory optimisation tasks. We introduce a novel \gls{RL} training framework tailored to \glspl{gcnet}, incorporating decoupled action and control frequencies alongside reward redistribution strategies to stabilise training and to provide a fair comparison. Our results show that \gls{BC}-trained \glspl{gcnet} excel at closely replicating expert policy behaviour, and thus the optimal control structure of a deterministic environment, but can be negatively constrained by the quality and coverage of the training dataset. In contrast \gls{RL}-trained \glspl{gcnet}, beyond demonstrating a superior adaptability to stochastic conditions, can also discover solutions that improve upon suboptimal expert demonstrations, sometimes revealing globally optimal strategies that eluded the generation of training samples.

\end{abstract}



\section{\label{sec:Intro}Introduction}
\glsresetall

Low-thrust propulsions is an established propulsive technology for interplanetary and deep-space missions, as demonstrated by missions such as ESA SMART-1 mission ~\cite{Racca1999}, NASA’s Deep Space 1 \cite{RAYMAN1999381}, Dawn and GRAIL missions, JAXA’s Hayabusa 1 \& 2 and recently ESA's BepiColombo~\cite{Thomas2012,Kawaguchi2012,Tsuda2013,Benkhoff2010}. Spacecraft autonomy is a major barrier to increasing the scope, ambition, and affordability such missions. \Glspl{gcnet} are emerging as a promising neural model for enhancing onboard autonomy and seamlessly incorporating optimality principles onboard spacecraft \cite{sanchez-sanchez_real-time_2018, izzo2023}, providing an alternative to conventional \gls{MPC} schemes by leveraging advancements in machine learning. \glspl{gcnet} are small, feed-forward \glspl{ANN} mapping the current state of a spacecraft to the corresponding optimal control action in a single inference thus offering an end-to-end differentiable representation of the entire spacecraft guidance and control architecture.



Two principal training philosophies have been so far employed for \glspl{gcnet}: \gls{BC} and \gls{RL}. \gls{BC} is a form of imitation learning that frames the \gls{GC} training task as supervised learning. Given a dataset of expert spacecraft trajectories (typically generated through numerical solutions to optimal control problems based on Bellman or Pontryagin principles), the \gls{gcnet} is trained to replicate the corresponding observation-action pairs. This enables the use of well-established supervised learning techniques to encode optimal guidance strategies directly into a neural policy~\cite{foster_is_2024}. In contrast, \gls{RL} trains a \gls{gcnet} through direct interaction with a simulated space environment, guided solely by scalar rewards rather than expert demonstrations. Here, the objective is to learn a policy that maximizes the expected cumulative reward over a trajectory, allowing the \gls{gcnet} to discover novel or improved guidance strategies autonomously. Unlike \gls{RL}, \gls{BC} lacks reward feedback and instead aims to mimic expert behaviour as closely as possible under the assumption of an unobserved reward structure~\cite{foster_is_2024}. This distinction is particularly relevant in space applications, where the trade-off between leveraging known optimal solutions as expert demonstrations and enabling autonomous adaptability becomes critical in real-world deployment scenarios.

In the context of spacecraft, \gls{BC} has been successfully used to train \glspl{gcnet} to control a spacecraft during a fuel-optimal orbit transfers~\cite{izzo2021} and rendezvous~\cite{evans2025fuel}, time-optimal low-thrust transfers~\cite{cheng2019, IzzoOriger2023_GCNETTime}, landing problems~\cite{mulekar2023metric, sanchez-sanchez_real-time_2018, CHENG2020_asteroid, origer_2024_SIREN} and hypersonic reentry~\cite{shi2020hypersonic}.

The use of \gls{RL} in decision-making systems has produced exciting results over the past decade in a diverse range of applications from robotics to self-driving cars, \gls{uav} and now spacecraft \cite{Sutton1998, Levine2013}. The allure of \gls{RL}-based algorithms for spacecraft guidance is their: (i) performance in unfamiliar environments \cite{Gaudet2019}, (ii) potential for creative/un-intuitive solutions \cite{silverMasteringGameGo2017}, (iii) similarities with optimal control theory \cite{Sutton1998}, (iv) track record of practical success \cite{mahmoodBenchmarkingReinforcementLearning2018, elsallabDeepReinforcementLearning2017, rodriguez-ramosDeepReinforcementLearning2019}, and (v) ability to generate optimal control policies \cite{nakamura-zimmererAdaptiveDeepLearning2021}. There is growing interest in applying \gls{RL} in astrodynamics \cite{Izzo2018a}, from periodic orbit transfers \cite{Miller2019, LaFarge2020, yanagidaExplorationLongTimeofFlight2020, Sullivan2020, federici2021autonomous} to station-keeping \cite{Bosanac2021_AAS}, rendezvous \cite{Scorsoglio2018, federiciDeepLearningTechniques2021}, landing \cite{Furfaro2020, GAUDET2020_landing}, interplanetary transfers~\cite{Zavoli2021, hu2023_denseRL}, solar-sail trajectories~\cite{bianchi_robust_2025} and even many-revolution transfers~\cite{Kwon2021, holt2021, holt2024reinforced}.

Some would argue \gls{BC} is preferable to \gls{RL} because it removes the need for exploration, leading to empirically reduced sample complexity and often much more stable training~\cite{foster_is_2024}. There are many studies demonstrating \gls{RL} algorithms are a good choice when the available data is either random or highly suboptimal~\cite{kumar_when_2022}. In the \gls{uav} community \gls{RL} has gained incredible success, surpassing the performance of human drone racing champions~\cite{kaufmann2023champion} and recently winning global drone racing tournaments (e.g. A2RL Grand Challenge 2025) with an approach based on \glspl{gcnet} \cite{Ferede2024_rldrone}. This serves as a compelling testament to \gls{RL}’s remarkable performance in the context of drone racing, where the availability of a dense reward function and uncertainties in the dynamics make the \gls{RL} approach efficient~\cite{izzo2023}, surpassing the performance of \gls{BC}~\cite{Ferede2024_bcdrone}. Spacecraft, however, operate in a rather different environment to drones, comparatively free of major disturbances, well modelled dynamics, and one in which optimality is of paramount importance. Thus in the context of spacecraft \glspl{gcnet}, it is still unclear when to prefer \gls{RL} over \gls{BC}. 



This paper presents a comprehensive comparison of \gls{BC} and \gls{RL} for training \glspl{gcnet}. Four different spacecraft transfer scenarios are considered, encompassing inertial and rotating reference frames, time- and mass-optimal transfers, spacecraft with high- and low-control authority, and different target event functions. A similarly broad selection of problems with relatively unchanged setups has not previously been considered in the literature, demonstrating both the versatility of the trained \glspl{gcnet} and allowing a more general reflection of \gls{BC} and \gls{RL} for spacecraft transfers. Whilst previous work by the authors has lead to significant improvements to the \gls{BC} framework \cite{izzo2021, IzzoOriger2023_GCNETTime, origer_2024_SIREN}, this paper also presents two notable additions to the \gls{RL} framework. A reward redistribution is introduced to aid with sparse terminal rewards, a problem that often plagues the use of \gls{RL} in astrodynamics~\cite{izzo2023, hu2023_denseRL}. This also ensures the same \gls{RL} approach works well for time-optimal, time-fixed mass-optimal and time-free mass-optimal scenarios. In addition, the control represented by the \glspl{gcnet} is numerically integrated as a function of time inside the \gls{RL} update function, rather than assuming its value as a constant or a Dirac function (i.e. impulsive). This simple, and yet unusual, addition decouples the control frequencies from the action frequencies during training, eventually allowing larger steps between actions in episode rollout without loss of optimality. 
Even more importantly, it also extends the use of \gls{RL} from multi-impulse and zero-order hold implementations to generic time-varying representations, allowing a direct comparison with \gls{BC}-\glspl{gcnet} and optimal control solutions. We deliberately align the structure of the \gls{BC}- and \gls{RL}-\glspl{gcnet} as much as possible to ensure the comparison is consistent. 

The remainder of this paper is structured as follows. Section~\ref{sec:Problem} outlines problem setup including the dynamical models and the four transfer scenarios considered. Section~\ref{sec:Training Methodology} includes the \gls{NN} architecture for the \glspl{gcnet} and discusses the training frameworks for both the \gls{BC} and \gls{RL} approaches. The key elements are the expert examples used for the \gls{BC} and the reward functions for \gls{RL}. Results are presented in \ref{sec:Results}, including a comparison of the computational cost, the optimality and robustness to stochastic uncertainties for each of the four transfer scenarios. Conclusions are drawn in Section~\ref{sec:Conclusion}.

\section{\label{sec:Problem}Problem Setup}

In this paper we use multiple optimal control problems as test cases. We consider both interplanetary rendezvous and small-body landing scenarios, and a selection of time- and fuel-optimal problems in a rotating and inertial reference frames. Table~\ref{table:taxonomy} gives a high-level taxonomy of the scenarios considered. The dynamics are given in Section~\ref{sec:InterDyn} and then the specific parameters of these scenarios are given in Section~\ref{sec:TransferScenarios}.

\begin{table}[b]
\centering
\caption{Taxonomy of scenarios considered. Control authority indicates the ratio between the acceleration capable from the spacecraft control system and the gravitational acceleration at the initial condition.}\label{table:taxonomy}
\adjustbox{max width=\linewidth}{
\begin{tabular}{llccccccc}
\toprule
Scenario & Problem & Case & \makecell{Time\\Optimal} & \makecell{Fuel\\Optimal} & \makecell{Inertial Reference \\Frame} & \makecell{Rotating Reference\\Frame} & \makecell{Control \\Authority} & \makecell{Event} \\
\midrule
Interplanetary Rendezvous & GTOC 11    & A & \checkmark & & & \checkmark & 0.0098 & \acrshort{SOI}\\
 & Earth-Mars & B & & \checkmark & \checkmark & & 0.0843 & \acrshort{SOI}\\
Small-body landing & Psyche   & C & & \checkmark & & \checkmark & 0.0591 & \acrshort{NN} \\
 & 67P         & D & & \checkmark & & \checkmark & 2.5172 & \acrshort{NN} \\
\bottomrule
\end{tabular}
}
\end{table}

\subsection{\label{sec:InterDyn}Dynamics}


\subsubsection{\label{sec:inertialframe}Inertial Frame}

In an inertial reference frame $\mathcal{F_I} = [\hat{\mathbf x}, \hat{\mathbf y}, \hat{\mathbf z}]$, the equations of motion can be written as:
\begin{equation}
\label{eq:dyn_inertial}
\left\{ 
\begin{array}{l}
    \dot{\bm{r}} = \bm{v}  \\
    \dot{\bm{v}} = -\frac{\mu}{r^3}\bm{r} + \frac{T_{\text{max}}}{m} \alpha \bm{u} \\
    \dot{m} = - \frac{T_{\text{max}}}{I_{sp} g_0}\alpha
\end{array}
\right.\, .
\end{equation}

The state vector $\mathbf{x}$ consists of the position $\mathbf{r}=[x,y,z]$, velocity $\mathbf{v}=[v_x,v_y,v_z]$, both expressed in the inertial frame $\mathcal{F_I}$, and spacecraft mass $m$. Here $r = \sqrt{x^2+y^2+z^2}$ and $\mu$ denotes the gravitational constant of the central body. 

The system is controlled by the thrust direction, represented by the unit vector $\mathbf{\hat{u}} = [u_x, u_y, u_z]$, and throttle factor $\alpha\in[0,1]$. The maximum thrust magnitude is denoted by $T_\text{max}$. The goal of the control problem is to determine a (piecewise continuous) function for $\mathbf{\hat{u}}(t)$ and time-of-flight $t_f$, where $t \in [t_0,t_f]$, so that, following the dynamics described by Eq.~\eqref{eq:dyn_inertial}, the state is steered from any initial state $\mathbf{r}_0$, $\mathbf{v}_0$, $m_0$ to a desired target state $\mathbf{r}_t$, $\mathbf{v}_t$. 

\subsubsection{\label{sec:rotatingframe}Rotating-frame}

In some cases of interest, we introduce a rotating frame $\mathcal{F_R} = [\hat{\mathbf i}, \hat{\mathbf j}, \hat{\mathbf k}]$ of angular velocity $\boldsymbol \Omega= \Omega \mathbf{\hat{k}}$, such that the target state, $\mathbf{r}_t$, $\mathbf{v}_t$, remains stationary within $\mathcal{F_R}$~\cite{IzzoOriger2023_GCNETTime, origer2024_issfd_neuralode}.

In an interplanetary rendezvous, if the target state, $\mathbf{r}_t$, $\mathbf{v}_t$, is in a circular orbit of radius $r_t$, then $\boldsymbol \Omega=\sqrt{\mu/r_t^3} \hat {\mathbf k}$. For the small-body pinpoint landing scenarios, the rotation rate $\boldsymbol \Omega$ is given by the body's rotation. In both cases, the position of the target state $\mathbf{r}_t=r_t\hat{\mathbf i}$ remains stationary in $\mathcal{F_R}$. The equations of motion in this rotating reference frame $\mathcal{F_R}$ are:

\begin{equation}
\label{eq:dyn_rotating}
\left\{ 
\begin{array}{l}
    \dot{\bm{r}} = \bm{v}  \\
    \dot{\bm{v}} = -\frac{\mu}{r^3}\bm{r} + 2 \Omega \times \dot{\bm{r}} + \Omega \times (\Omega \times \bm{r}) + \frac{T_{\text{max}}}{m} \alpha \bm{u} \\
    \dot{m} = - \frac{T_{\text{max}}}{I_{sp} g_0}\alpha
\end{array}
\right.\, .
\end{equation}

Here, the state vector $\mathbf{x}$ consists of the position $\mathbf{r}=[x,y,z]$ and velocity $\mathbf{v}=[v_x,v_y,v_z]$, both expressed in the rotating frame $\mathcal{F_R}$, and spacecraft mass $m$. Again, $r = \sqrt{x^2+y^2+z^2}$ and $\mu$ denotes the gravitational constant of the central body. The system is controlled by the thrust direction, represented by the unit vector $\mathbf{\hat{u}} = [u_x, u_y, u_z]$, and thrust magnitude $\frac{T_{\text{max}}}{m}$ and throttle factor $\alpha\in[0,1]$. The goal remains to determine a (piecewise continuous) function for $\mathbf{\hat{u}}(t)$ and time-of-flight $t_f$, where $t \in [t_0,t_f]$, so that, following the dynamics described by Eq.~\eqref{eq:dyn_rotating}, the state is steered from any initial state $\mathbf{r}_0$, $\mathbf{v}_0$, $m_0$, to a desired target state $\mathbf{r}_t=r_t\hat{\mathbf i}$, $\mathbf{v}_t=0$.

\subsection{\label{sec:TransferScenarios}Transfer scenarios}

\subsubsection{\label{Interplanetary}Interplanetary Rendezvous}
\begin{enumerate}[label=\Alph*)]
    \item \textit{GTOC 11}: Time-optimal transfer with constant acceleration to a circular orbit (in rotating reference frame $\mathcal{F_R}$) \cite{IzzoOriger2023_GCNETTime}. The dynamics are given in Eq.~\eqref{eq:dyn_rotating} where the specific impulse is infinite and thus the mass equation is removed.

    \item \textit{Earth-Mars:} Fuel-optimal transfer to an elliptical orbit (in inertial reference frame $\mathcal{F_I}$) \cite{Zavoli2021}. The dynamics are given in Eq.~\eqref{eq:dyn_inertial}.
\end{enumerate}

\noindent Parameters which remain constant across the simulations can be found in Table~\ref{table:TestCasePars}. The test-case-specific parameters are given in Table~\ref{table:TestCase}.

\begin{table}
\centering
\caption{Interplanetary Rendezvous Test Case Parameters}
\label{table:TestCasePars}
\adjustbox{max width=\linewidth}{
\begin{tabular}{ll cc} 
    \toprule
    Name & Variable & GTOC 11 & Earth-Mars\\ [0.5ex] 
    \midrule
    Gravitational acceleration at sea level & $g_0$ & \multicolumn{2}{c}{$9.80665$ m/s}\\
    Gravitational parameter (Sun) & $\mu_S$ & \multicolumn{2}{c}{ $1.32712440018e20$ m$^3$/s$^2$ } \\
    Astronomical Unit & $L$ & \multicolumn{2}{c}{$149597870691.0$ m} \\
    \midrule
    Rotation Rate & $\Omega$ & $1.34e-7$ rad/s & $0$ rad/s\\
    Thrust & $T_{\text{max}}$ & $100$ mN & $500$ mN \\
    Specific Impulse & $I_{sp}$ & $\infty$ & $2000$ s \\
    Spacecraft Mass & $m_0$ & $1000$ kg  & $1000$ kg \\
    Maximum time-of-flight & $t_f$ & free & $348.79$ days \\
    Position Convergence & $c_r$ & 924,000 km & 577,000 km \\   
    Velocity Convergence & $c_v$ & 500 m/s & 1000 m/s\\
    \bottomrule
\end{tabular}
}
\end{table}

\begin{table}[h!]
\centering
\caption{Interplanetary Rendezvous Test Cases (inertial reference frame)}
\label{table:TestCase}
\adjustbox{max width=\linewidth}{
\begin{tabular}{ll cccccc} 
    \toprule
    Case & Objective & \multicolumn{4}{c}{Initial Conditions [AU, AU/yr]}\\
    \midrule
    A & Spacecraft & $\bm{x}$ & -1.18743886 & -3.05783963 & 0.3569407\\
    & & $\bm{v}$ & 0.44567591 & -0.18673354 & 0.02152004 \\
    & Target & $\bm{x}$ &1.3 & 0.0 & 0.0 \\
    & & $\bm{v}$ & 0.0 & 0.8770580193070292 & 0.0 \\
    \midrule
    B & Spacecraft & $\bm{x}$ & -0.9405193559915066 & -0.3450211407528088 & 6.550895380217187e-06 \\
    &  & $\bm{v}$ & 0.3281752940382571 & -0.9427090989497672 & 1.4563521504202196e-05\\
    & Target & $\bm{x}$ & 0.6049580035267025 & -1.2735875745977223 & -0.041541980167412354 \\
    &  & $\bm{v}$ & 0.7655476388773976 & 0.4187780440110384 & -0.010029635695970087 \\
    \bottomrule
\end{tabular}
}
\end{table}

\subsubsection{\label{SmallBodyLanding}Small-body Landing}
\begin{enumerate}[label=\Alph*)]
    \setcounter{enumi}{2} 
    \item \textit{Psyche:} Fuel-optimal landing on asteroid Psyche (in rotating reference frame $\mathcal{F_R}$). The dynamics is given in \eqref{eq:dyn_rotating}.
    \item \textit{67P:} Fuel-optimal landing on comet Churyumov-Gerasimenko 67P (in rotating reference frame $\mathcal{F_R}$, high control authority) \cite{izzo2025high}. The dynamics is given in Eq.~\eqref{eq:dyn_rotating}.
\end{enumerate}

\noindent Parameters which remain constant across the simulations can be found in Table~\ref{table:TestCasePars_Asteroid}. The test-case-specific parameters are give in Table~\ref{table:TestCase_Asteroid}.

\begin{table}[h!]
\centering
\caption{Small-body Landing Test Case Parameters}
\label{table:TestCasePars_Asteroid}
\adjustbox{max width=\linewidth}{
\begin{tabular}{ll cc} 
    \toprule
    Name & Variable & Psyche & 67P \\ [0.5ex] 
    \midrule
    Gravitational acceleration at sea level & $g_0$ & \multicolumn{2}{c}{$9.8$ m/s}\\
    \midrule
    Gravitational parameter (Small Body) & $\mu$ & $1.530348200e9$ m$^3$/s$^2$ & $6.674e2$ m$^3$/s$^2$ \\
    Rotation Rate & $\Omega$ & $4.159558822-4$ rad/s & $1.367705706e-4$ rad/s\\
    Thrust & $T_{\text{max}}$ &  $80$ mN &  $10.5$ mN \\
    Specific Impulse & $I_{sp}$ &  $200$ s &  $100$ s \\
    Spacecraft Mass & $m_0$ & $353.405305$ kg & $100$ kg \\
    Asteroid Event Altitude & $c_{NN}$ & 1 km & 0 m \\   
    Position Convergence & $c_r$ & 2 km & 5 m \\   
    Velocity Convergence & $c_v$ & 25 m/s & 0.05 m/s \\
    \bottomrule
\end{tabular}
}
\end{table}

\begin{table}[h!]
\centering
\caption{Small-body Landing Test Cases (rotating reference frame)}
\label{table:TestCase_Asteroid}
\adjustbox{max width=\linewidth}{
\begin{tabular}{lll cccccc} 
    \toprule
    Case & Body & Object & \multicolumn{4}{c}{Initial Conditions [m, m/s]}\\
    \midrule
    C & Psyche & Spacecraft & $\bm{x}$ & 180000.0 & 10000.0 & 0.0 \\
    & & & $\bm{v}$ & 25.0 & -25.0 & 20.0 \\
    & & Target & $\bm{x}$ & 122241.295 & -4889.878 & -1638.576 \\
    & & & $\bm{v}$ & 0.0 & 0.0 & 0.0 \\
    \midrule
    D & 67P  & Spacecraft & $\bm{x}$ &-7963.0 & -437.0 & 3452.0  \\
    & & & $\bm{v}$ & -0.4285 & 1.312 & -0.6158  \\
    & & Target & $\bm{x}$ &2317.93 & -178.89 & 71.547 \\
    & & & $\bm{v}$ & 0.0 & 0.0 & 0.0 \\
    \bottomrule
\end{tabular}
}
\end{table}

\subsubsection{\label{sec:Events}Convergence Criteria: Events}

In order to evaluate the quality of a \gls{gcnet}, we need to define a convergence criteria (in both position and velocity) around the target state to terminate the trajectory. Let $c_r$ and $c_v$ be the convergence radii for position and velocity. A \gls{gcnet} trajectory is considered to have converged if both $e_{r} = ||\bm{r}-\bm{r}_t||<c_r$ and $e_{v} =||\bm{v}-\bm{v}_t||<c_v$ simultaneously.

In practice, the \glspl{gcnet} are numerically integrated using a taylor-adaptive integrator (in \textit{heyoka.py} \cite{Biscani_Heyoka}) with a position \textit{event}-manifold to terminate the integration, using reliable event-detection machinery~\cite{biscani_reliable_2022}. For the interplanetary rendezvous, the target body's \gls{SOI} can be used to define the position threshold, $c_r$ and acting as the \textit{event}-manifold where the integration can be terminated. A suitable velocity threshold $c_v$ can then be used to assess if the trajectory has converged to the target. If used onboard a spacecraft, a different guidance and control scheme could then be deployed for the final approach. For the small-body landing, the complex three-dimensional shape of the body needs to be considered. This is done by training a small \gls{NN} to represent a boundary defined by a given altitude above the asteroid’s surface, $c_{NN}$. For more detail see \cite{origer2024_certifying_IAC, izzo2025high}. Once the trajectory has reached this \gls{NN}-event the integration is terminated. The trajectory is only considered converged to the target state if it is then within a sphere of radius $c_r$ in position and $c_v$ in velocity. Note, this can be a source of confusion in the remainder of the manuscript. The \gls{NN}-event is a separate network from the \gls{gcnet} - the two are not linked in any way. The \gls{NN}-event is used to ensure the terminal condition is differentiable, enabling the use of reliable event-detection machinery~\cite{biscani_reliable_2022}.

\section{\label{sec:Training Methodology}Training Methodology}

\subsection{\label{sec:Network Architecture}Network Architecture}

As seen in Section~\ref{sec:InterDyn}, if the dynamics are autonomous and the control $\bm{u}=\alpha \hat{\bm{u}}$ is a function of the state $\bm{x}$, the dynamics can be written as
\begin{equation}
\label{eq:dyn_auto}
\begin{array}{l}
    \dot{\bm{x}} = f(\bm{x}) + g(\bm{x})\bm{u}(\bm{x}).\\
\end{array}
\end{equation}
For a \gls{gcnet}, this feedback control law $\bm{u}(\bm{x})$ is given by a \gls{NN}, $\bm{u}_{NN}(\bm{x})=\mathcal{NN}_{\bm{\theta}}(\bm{x})$. The architectures used in this paper are discussed here.

\begin{figure*}[htb]
  \centering
  \includegraphics[width=0.48\textwidth]{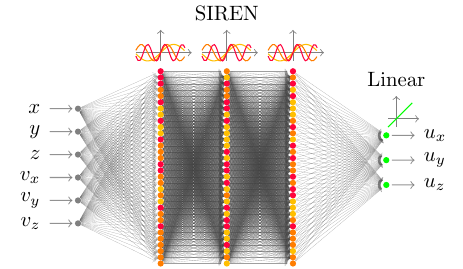}
  \hfill
  \includegraphics[width=0.48\textwidth]{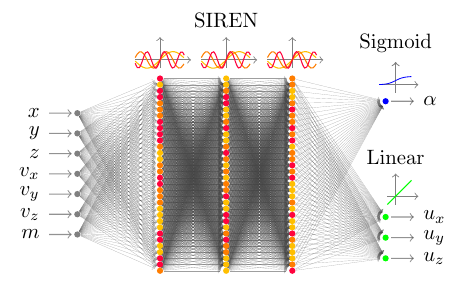}
  \hfill
 \caption{\gls{gcnet} architectures using SIREN \cite{sitzmann2019siren} and Linear activation functions for time-optimal (left) and fuel-optimal (right) transfers}
  \label{fig:ffnn_siren}
\end{figure*}

\begin{figure*}[htb]
  \centering
  \includegraphics[width=0.48\textwidth]{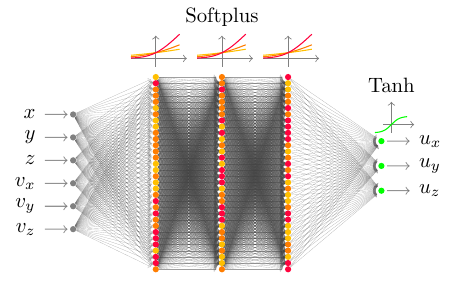}
  \hfill
  \includegraphics[width=0.48\textwidth]{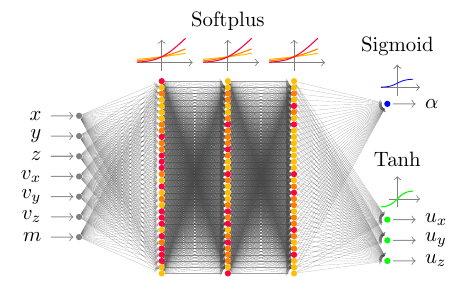}
  \hfill
  \caption{\gls{gcnet} architectures using Softplus and Tanh activation functions for time-optimal (left) and fuel-optimal (right) transfers}
  \label{fig:ffnn_softplus}
\end{figure*}

Each \gls{gcnet} has three hidden layers with $32$ neurons each, which amounts to $2435$ and $2500$ parameters for the time- and fuel-optimal scenarios. This is lower than most other works, such as the $6196$ used in \cite{Zavoli2021}, $120,000$ used in \cite{izzo2021} and $34,307$ initially required in \cite{origer2024_issfd_neuralode}. Whilst we aim to keep as much of the architecture the same across the two training approaches to ensure a consistent comparisons, it is necessary to change the activation functions. In \cite{origer_2024_SIREN} the authors found that when the \glspl{gcnet} are trained via \gls{BC}, using a periodic activation function for the hidden layer results in much more accurate networks. These findings were inspired by the work of Sitzmann et al. \cite{sitzmann2019siren}, who came up with \glspl{SIREN} which showed very impressive approximation power for image and video reconstruction, as well as complex boundary value problems, surpassing more common activation functions. However, similar performance benefits were not observed for \gls{RL}-\glspl{gcnet}. In fact, \cite{mavor-parker_frequency_2024} explore the use of periodic activation functions for \gls{RL} and find there is still a generalisation gap to be closed between Fourier representations and ReLU representations. As such, we stick with traditional activation functions for the \gls{RL}-\glspl{gcnet}, using Softplus activation functions instead of ReLUs to ensure differentiability and enable the use of a taylor-adaptive integrator. Figures~\ref{fig:ffnn_siren} and~\ref{fig:ffnn_softplus} show the \gls{BC} and \gls{RL} \gls{gcnet} architectures respectively.


\subsection{\label{sec:Behavioural Cloning}Behavioural Cloning}

When training \glspl{gcnet} on datasets of optimal trajectories using \gls{BC} we make use of a few recent results that improve the overall training pipeline. Since all optimal control problems here considered can be solved with Pontryagin's Maximum principle (interplanetary rendezvous and small-body landing), we leverage a technique called the \gls{bgoe} \cite{izzo2021,IzzoOriger2023_GCNETTime}. This allows us to generate very efficiently hundreds of thousands of optimal trajectories by perturbing the final co-states of one single nominal solution. 
Once these trajectories are obtained they are sampled in 100 points. All these state-action pairs are then be used as features and labels respectively in the \gls{BC} pipeline.
In all the cases we use a 80/20 split for training and validation data, the Adam optimiser \cite{kingma2014adam} and a scheduler that decreases the learning rate by 10\% whenever the loss fails to improve for 10 consecutive epochs.
The loss function for the time-optimal scenario is: $\mathcal L= 1-\frac{\mathbf{\hat{u}}^*\cdot \mathbf{\hat{u}}_{NN}}{\mathbf{\vert\hat{t}}^*\vert\vert \mathbf{\hat{u}}_{NN}\vert}$, hence the \gls{gcnet} learns to minimise the cosine similarity between the estimated thrust direction $\mathbf{\hat{u}}_{NN}$ and the ground truth $\mathbf{\hat{u}}^*$. For the fuel-optimal scenarios we add an additional term which penalises the mean squared error between the estimated throttle $\alpha_{NN}$ and the ground truth $\alpha^*$:  $\mathcal L= \text{MSE}(\alpha_{NN},\alpha^*) + 1-\frac{\mathbf{\hat{u}}^*\cdot \mathbf{\hat{u}}_{NN}}{\mathbf{\vert\hat{t}}^*\vert\vert \mathbf{\hat{u}}_{NN}\vert}$. 

\begin{figure*}[htb]
  \centering
  \includegraphics[width=\textwidth]{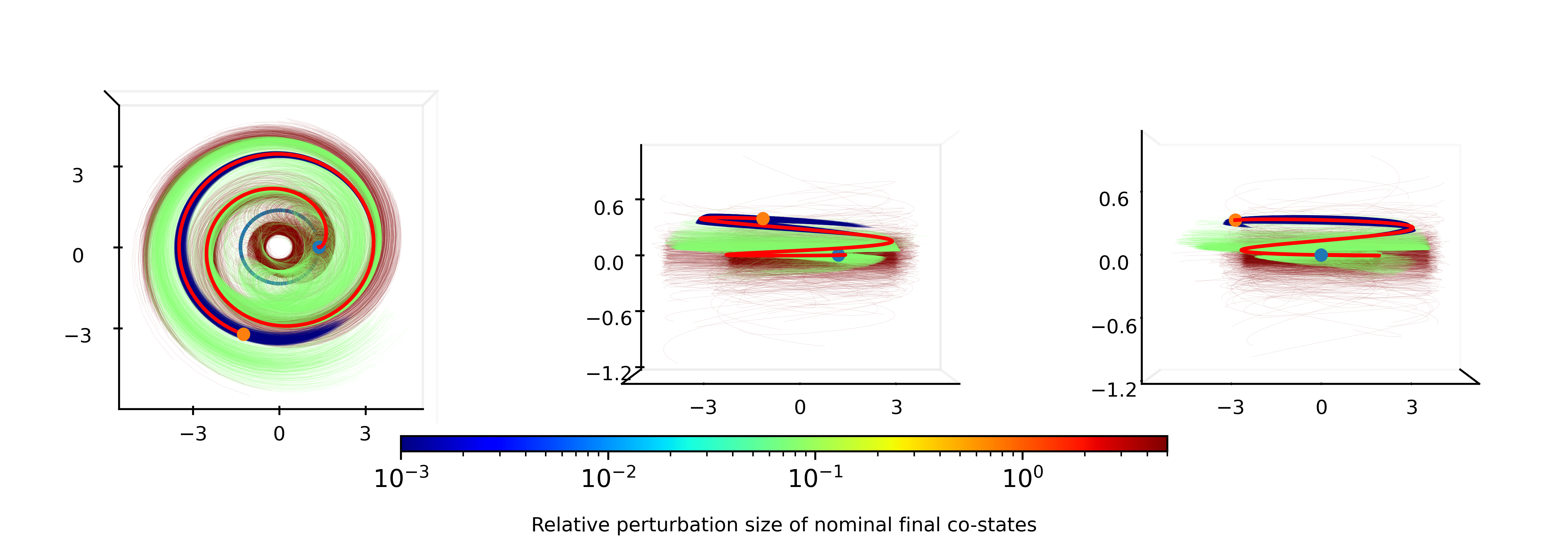}
  \caption{Trajectories generated using BGOE for case A (\textit{GTOC 11}), seen in the rotating frame. Astronomical units used. From left to right: YX, ZY and ZX projections. Nominal trajectory shown in red.}
  \label{fig:bgoe_caseA}
\end{figure*}

\begin{figure*}[htb]
  \centering
  \includegraphics[width=\textwidth]{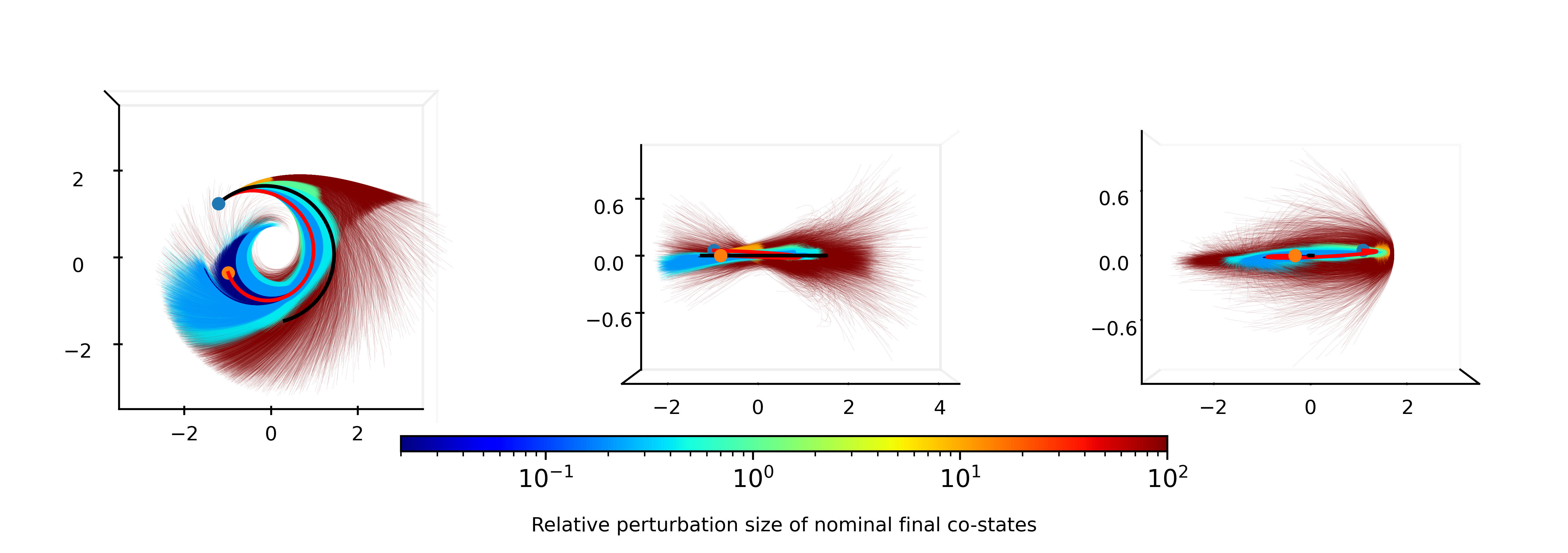}
  \caption{Trajectories generated using BGOE for case B (\textit{Earth-Mars}), seen in the inertial frame. Astronomical units used. From left to right: YX, ZY and ZX projections. Nominal trajectory shown in red.}
  \label{fig:bgoe_caseB}
\end{figure*}

\begin{figure*}[htb]
  \centering
  \includegraphics[width=\textwidth]{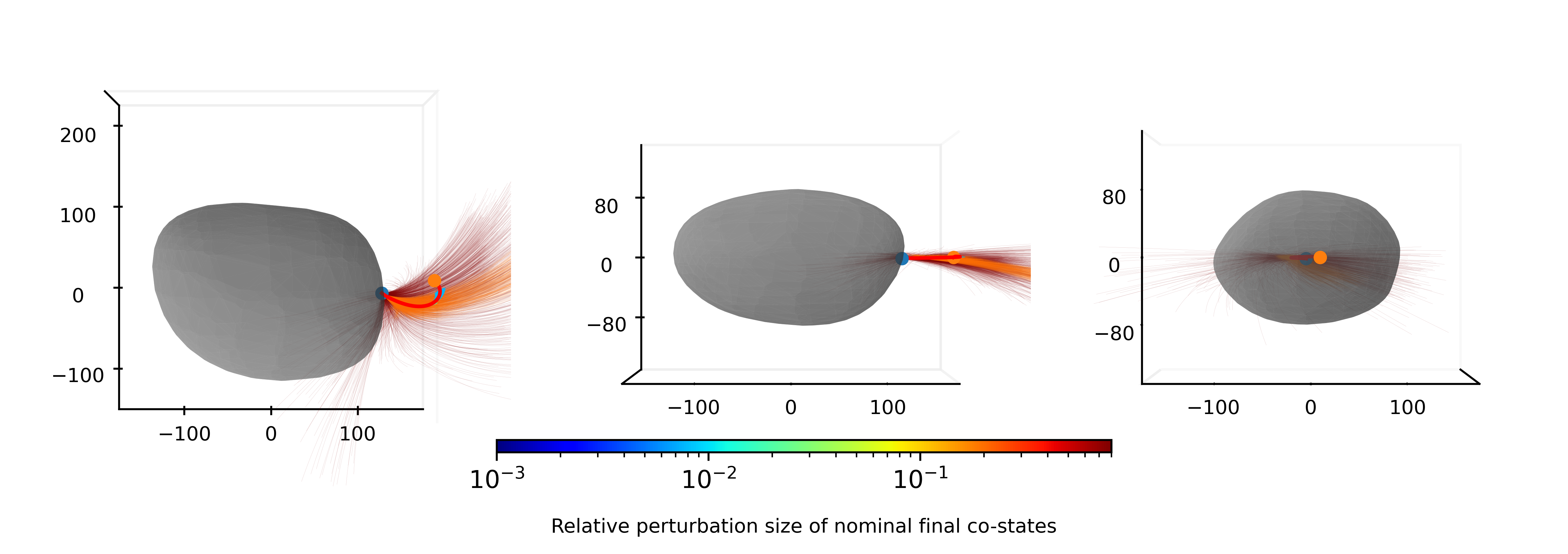}
  \caption{Trajectories generated using BGOE for Psyche, seen in the rotating frame. Axis units in kilometres. From left to right: YX, ZY and ZX projections. Nominal trajectory shown in red.}
  \label{fig:bgoe_casePsyche}
\end{figure*}

\begin{figure*}[htb]
  \centering
  \includegraphics[width=\textwidth]{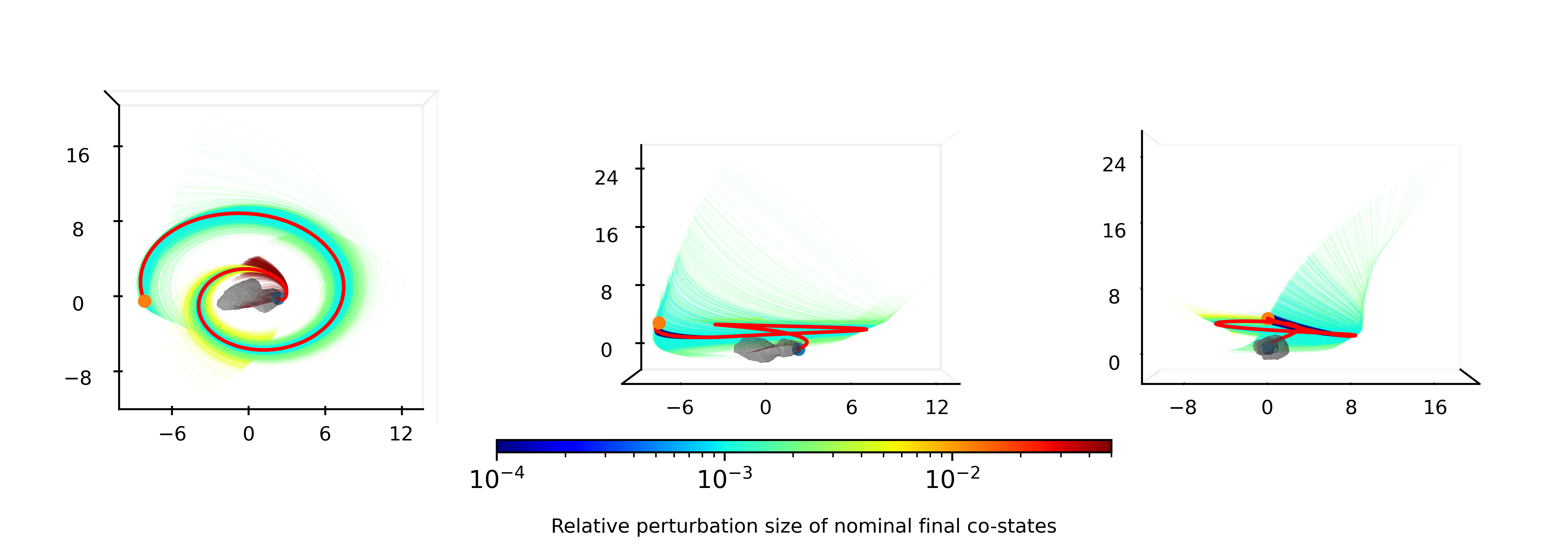}
  \caption{Trajectories generated using BGOE for 67P, seen in the rotating frame. Axis units in kilometres. From left to right: YX, ZY and ZX projections. Nominal trajectory shown in red.}
  \label{fig:bgoe_case67P}
\end{figure*}

The training databases for both interplanetary and small-body landing scenarios are made up of smaller bundles each generated with varying costate perturbation magnitudes and time-of-flights. These different costate perturbations are needed to address distribution shift--a common challenge in \gls{BC}. Interested readers are referred to \cite{moreno-torres_unifying_2012, laskey_dart_nodate, shin2024distillingprivilegedinformationdubins} for further discussion. For the interplanetary case A (\textit{GTOC 11}), $4$ bundles of $100,000$ trajectories are generated for different costate perturbation magnitudes, whilst $7$ bundles of $50,000$ are used for case B (\textit{Earth-Mars}). These are shown in Figs.~\ref{fig:bgoe_caseA} and ~\ref{fig:bgoe_caseB}. For the small-body landings, Psyche also has $7$ bundles of $50,000$ trajectories, whilst 67P uses $6$ bundles of $40,000$. These are shown in Figs.~\ref{fig:bgoe_casePsyche} and ~\ref{fig:bgoe_case67P}. We use the following hyperparameters: 4096 as the batch size (training and validation), $5\text{e-5}$ as the learning rate, weight decay values of $2.5\text{e-5}$, $2.5\text{e-5}$, and $0.0$ respectively, and training epochs of 500, 500, and 200 respectively.

\subsection{\label{sec:Reinforcement Learning}Reinforcement Learning}


\gls{RL} problems are usually posed within a \gls{MDP}, as sequence of state-action pairs $\bm{x}_i$ and $\bm{a}_i$. The agent (in our case the spacecraft) interacts with the environment (the dynamics) using a parametrised policy $\pi_\theta (\bm{a}|\bm{x})$ (the actor network, which is represented by the \gls{gcnet}). This determines the action taken give the current state $\bm{a} \sim \pi_\theta (\bm{a}|\bm{x})$. As the agent interacts with the \gls{MDP} it collects rewards $r_i=r(\bm{x}_i, \bm{a}_i)$ based on the actions taken. The agent's goal is to obtain a policy that maximises the cumulative reward (or if you prefer, minimises the cumulative cost) from the start state to the end state.


In this work an \gls{RL} update strategy based on \gls{PPO} \cite{schulman2017proximal} is used. This is an actor-critic on-policy algorithm which clips the objective function to remove incentives for the new policy to get too far away from the old policy. In other words it ensures the update size is within a trusted region, attempting to prevent accidentally bad updates. The results presented in the remainder of this paper were obtained using the \gls{PPO} implementation from the Stable Baselines3 library \cite{stable-baselines3}. \Gls{PPO} is chosen due to its frequent use in astrodynamics, robustness to hyperparameters and stable learning curves \cite{Federici2023}. 

\gls{PPO} uses a stochastic policy during training and a deterministic one during evaluation. The actions are sampled from a normal distribution with mean $\bar{\bm{a}}=[\bar{\alpha}, \bar{u}_x, \bar{u}_y, \bar{u}_z]$ and standard deviations $\bm{\sigma}=[\sigma_\alpha, \sigma_x, \sigma_y, \sigma_z]$.  The \glspl{gcnet} shown in Fig.~\ref{fig:ffnn_softplus} therefore have an additional set of weights (for the additional outputs $\bm{\sigma}$) that are updated during training. However, when evaluating the \gls{gcnet}, a deterministic setup is used where the actions correspond to their mean values $\bar{\bm{a}}=[\bar{\alpha}, \bar{u}_x, \bar{u}_y, \bar{u}_z]$. At the start of training, the agent will take actions based on an untrained policy, and the stochasticity enables it to explore the environment. As it gets more confident in its actions and seeks to optimise the objective, it will reduce the stochastic exploration by reducing $\bm{\sigma}$. 

Conventional \gls{RL} trains by sampling the actions at step $i$ and then keeping them constant to step $i+1$, before sampling them again. For a \gls{gcnet}, this action corresponds to a control vector $\bm{u}_i\sim\mathcal{N}(\bar{\bm{u}}_i,\sigma_i)=\bar{\bm{u}}_i+\delta \bm{u}_i$, which is then held in a \gls{ZOH}. Thus the next state is computed as:
\begin{equation}
\label{equ:IntegrateZOH}
\begin{array}{l}
    \bm{x}_{i+1} = \int_{t_i}^{t_{i+1}} f(\bm{x}) + g(\bm{x})\bm{u}_{NN}(\bm{x}(t_i)) dt\\
    \bm{x}_{i+1} = \int_{t_i}^{t_{i+1}} f(\bm{x}) + g(\bm{x})(\bar{\bm{u}}_i+\delta \bm{u}_i) dt\\
\end{array}
\end{equation}

\noindent In this work we present a modification to this convention, to obtain a continuous representation of the control as:
\begin{equation}
\label{eq:IntegrateRHS}
\begin{array}{l}
    \bm{x}_{i+1} = \int_{t_i}^{t_{i+1}} f(\bm{x}) + g(\bm{x})\bm{u}_{NN}(\bm{x}) dt\\
    \bm{x}_{i+1} = \int_{t_i}^{t_{i+1}} f(\bm{x}) + g(\bm{x})(\bar{\bm{u}}_{NN}(\bm{x})+\delta \bm{u}_i) dt\\
\end{array}
\end{equation}
where $\bar{\bm{u}}_{NN}(\bm{x})$ is obtained by numerically integrating the \gls{gcnet} as a function of time inside the integration routine, and thereby inferring it at the frequency of the integrator. This is done using the \textit{heyoka.py} toolbox \cite{Biscani_Heyoka} and allows us to decouple the control frequencies from the step/action frequencies, eventually allowing larger steps between actions in episode rollout without loss of optimality. This was essential for solving case A in particular. Although not presented here, this structure means the same \gls{RL} framework can be used to generate both continuous-thrust, \gls{ZOH} thrust and multi-impulse \gls{gcnet} policies.

\subsubsection{\label{sec:RL_LossTime}Reward Function: Time-optimal}

The reward function needs to have two components: encourage convergence to the target, $r_x$ and optimise some objective (e.g. time of flight or propellant mass) $r_o$. Once the \gls{gcnet} trajectory has terminated, we could start with a reward function that tries to minimise the final position $e_{r} = ||\bm{r}-\bm{r}_t||$ and velocity errors $e_{v} =||\bm{v}-\bm{v}_t||$ independently:
\begin{equation}
    r_x = -e_{r} - e_{v}.
\end{equation}
From experience this proves problematic as it often tries to get $e_{v} \sim 0$ without driving $e_{r} \rightarrow 0$, which means we aren't very close to the target. This creates local minima. In addition, the scaling of $e_{r}$ and $e_{v}$ is quite arbitrary. In turn, we can divide these by the convergence radii and we can use a logarithmic scale to prevent the cost from growing too much when far away from the target. Similar reasoning lead to the exponential terminal reward was used in \cite{hu2023_denseRL}. This leads to a cost function of this form:
\begin{equation}
    r_x = -\log{\max{\left(\frac{e_{r}}{c_r}, 1\right)}} - \log{\max{\left(\frac{e_{v}}{c_v}, 1\right)}}.
\end{equation}
However, this still has the issue of a local minimum at $\frac{e_{v}}{c_v}\sim 1$ and $\frac{e_{r}}{c_r}>>1$. We notice we only need the velocity to go to zero when the position is close to converging. Hence, using a linear scaling, we can increase the effective size of $c_v$ depending on the position error, $c_v \gets c_v \frac{e_{r}}{c_r}$.
\begin{equation}\label{equ:Lx_time}
    r_x = -\log{\max{\left(\frac{e_{r}}{c_r}, 1\right)}} - \log{\max{\left(\frac{e_{v}}{c_v}\frac{c_r}{e_{r}}, 1\right)}}
\end{equation}

More complex functional forms might also work, but this linear relation helps to drive $\frac{e_{r}}{c_r}\sim 1$ first. Once $e_{r} = ||r-r_t||<c_r$ and $e_{v} =||v-v_t||<c_v$, $r_x = 0$, and we can start optimising the time-of-flight using:
\begin{equation}\label{equ:Lo_time}
    r_o = -\frac{t}{t_{f}}.
\end{equation}

\subsubsection{\label{sec:RL_LossMass}Reward Function: Fuel-optimal}

For the fuel-optimal transfers, a first step would be to modify the loss to represent the delta-v of the continuous-thrust arc, $r_o = -\Delta v_{LT} = I_{sp} g_0 \log{(m_f/m_0)}$. This works well when the event function is easy to reach (i.e. $r_x=0$). However, if the learning struggles to reach the event, then it needs to use more fuel to do so. This can result is a chattering during the learning process, as $r_x$ encourages more fuel usage, which in turn increases the magnitude of $r_o$. This constraint satisfaction issue is common in \gls{RL} training, and often leads to soft constraints in the cost function.

An alternative methodology is to rewrite the position and velocity constraints in terms of fuel-consumption. One way of doing this involves using a lambert arc to convert a position error to a $\Delta v$, which can, in turn, be converted to fuel consumption. An initial criticism might be that lambert arcs use impulsive $\Delta v$s, which conflicts with the continuous control approach considered in this work. Given the use of the events in this work, if the lambert arc can be consigned to inside this event, then the whole transfer is effectively done with the \gls{gcnet}. Thus the lambert arc would only be required to aid convergence during training and is ignored during validation and inference. 

We can update Eq.~\eqref{equ:Lx_time} with
\begin{equation}\label{equ:Lx_fuel}
    r_x = -\left(1+\log{\max\left(\frac{\Delta v_1}{\Delta v_1^{\text{max}}},1\right)}\right)\Delta v_1 - \left(1+\log{\max\left(\frac{\Delta v_2}{\Delta v_2^{\text{max}}},1\right)}\right)\Delta v_2.
\end{equation}
Here $\Delta v_1$ and $\Delta v_2$ represent the lambert arc $\Delta v$s. These are scaled by $\Delta v_1^{\text{max}}$ and $\Delta v_2^{\text{max}}$, which represent the maximum $\Delta v$ achievable by the spacecraft given its maximum thrust $T_{\text{max}}$ and engine efficiency $I_{sp}$. Namely, $\Delta v_i^{\text{max}} = (T_{\text{max}}/m_i) \Delta t_L$ where $m_i$ is the spacecraft mass at the start (1) or end (2) of the lambert arc. The unknown parameter is $\Delta t_L$, the duration of the Lambert arc. If we make $\Delta t_L$ very short, then the continuous-thrust part of the transfer (given by the \gls{gcnet}) is encouraged to get close to the target before the lambert arc is initiated. However, we observed that using a fixed value can be detrimental to the learning, and its best to consider a very small grid search on $\Delta t_L$ to encourage convergence. A suitable range of values can be considered from the convergence velocity $c_v$ and the duration it would take the continuous thrust of the spacecraft to accrue this $\Delta v$ (i.e. $\Delta t_L = \alpha_L \frac{c_v T_{\text{max}}}{m_i}$). Here $\alpha_L \in (0,1]$ acts as a scaling parameter to ensure $\Delta t_L$ is short enough such that the lamber arc is inside the event manifold. For this work we use $\alpha_L=0.1$.

A qualitative description of the overall reward function is it represents the total $\Delta v$ of both the continuous-thrust part $\Delta v_{LT}$ and the lambert arc $\Delta v_{1}+\Delta v_{2}$. However, the lambert arc  $\Delta v$s are scaled by the logarithmic terms and $\Delta v_i^{\text{max}}$ such that the end state of the continuous-thrust arc is as close to the target as possible. Indeed, using more fuel in the continuous-thrust arc will increase $\Delta v_i^{\text{max}}$ and thus lower the magntiude of $r_x$. We found this approach alleviated the chattering during learning and helped aid convergence, particularly for interplanetary case B (\textit{Earth-Mars}) and for the landing on 67P. Psyche was less affected because the \textit{event}-manifold is comparatively much larger. 

\subsubsection{\label{sec:RL_RedistReward}Reward Redistribution}

Many of the advantages of \gls{PPO}, and many \gls{RL} algorithms, are best harnessed when a state-dependent and thereby frequent reward function $r(x_i, a_i)$ can be provided. This poses an issue for spacecraft trajectory design. The quality of a trajectory or guidance law is often judged by time-of-flight, propellant mass consumed or terminal constraint accuracy. Each of these is best assessed on completion of an episode. The reward functions outlined in Section~\ref{sec:RL_LossTime} and~\ref{sec:RL_LossMass} are, as such, terminal rewards, and not state dependent ones. Such terminal rewards are also used in \cite{Zavoli2021}. In \cite{Miller2019}, amongst others, the state errors of spacecraft with respect to the target is used at each step (e.g. $r_x$ at each step). However, this is not suitable for multi-revolution problems, as noted by \cite{hu2023_denseRL}, and encourages a structure that might not be representative of the true optimal control solution.

For a sequence of states, the conventional discounted reward-to-go $R_i$ is used to redistribute terminal rewards, which is a discounted sum of the reward $r_i = r(x_i,a_i)$ at each remaining state along the sequence with discount factor $\gamma \in (0,1]$, and is written as:
\begin{equation}\label{equ:MDPcost}
\begin{aligned}
  R_i & = r_i + \gamma r_{i+1} + \gamma^2 r_{i+2} + ... = \sum_{j=i}^{\infty} \gamma^{j-i} r_j.
\end{aligned}
\end{equation}
This works well if the expected number of episodic steps is approximately the same, such as a time-fixed mass-optimal scenario considered in~\cite{Zavoli2021, hu2023_denseRL}. However, in order to generalise such that the same \gls{RL} approach works well for time-optimal, time-fixed mass-optimal and time-free mass-optimal, we propose an alternative, where the terminal reward can be assigned to any state along the trajectory and redistributed independent of how many steps were taken. 

This is explained in the following schematic:
\begin{equation}
\label{equ:redistributeReward_differentSteps}
\setlength{\arraycolsep}{4pt} 
\begin{array}{l@{\hskip 4pt}cccccccc}
    \textbf{Step:} & 1 & \dots & i & \dots & N & \dots & N + D \\ 
    \textbf{Time:} & t_1 & \dots & t_i & \dots & t_N & \dots & t_{N+D}\\
    \textbf{Time-step:} & \delta t_1 & \dots & \delta t_{i} & \dots & \delta t_{N} & \dots & \delta t_{N+D}\\
    \textbf{State:} & x_1 & \dots & x_i & \dots & x_N & \dots & \delta x_{N+D}\\
    \textbf{Action:} & a_1 & \dots & a_i & \dots & a_N & \dots & \delta a_{N+D} \\
    \textbf{Reward:} & 0 & \dots & 0 & \dots & 0 & \dots & r_x+r_o \\
    \textbf{Truncated Reward:} & 0 & \dots & 0 & \dots & r_x+r_o \\ 
    \textbf{Redistributed Reward:} & r_o\frac{\delta t_{1}}{t_N} & \dots & r_o\frac{\delta t_{i}}{t_N} & \dots & r_x \\
    \textbf{Returns:} & r_x+r_o & \dots & r_x+r_o\frac{(t_N-\sum_{j=0}^i\delta t_j)}{t_N} & \dots & r_x \\
\end{array}
\end{equation}
Here, $N+D$ represents the total number of steps sampled along a trajectory, $N$ steps are retained and the remaining $D$ are discarded. One advantage is you can propagate for $N+D$ steps and then use an alternative terminating factor, such as the closest approach to the target, to retroactively terminate the trajectory there, say after $N$ steps. This not only holds if each step has the same $\delta t$, but also if they have different $\delta t_i$, allowing us to vary the duration of each step. In addition, unlike in Eq.~\eqref{equ:MDPcost} were a different number of steps would lead to a different discount for the same trajectory, now the same trajectory can be divided into different numbers of steps and yet the discounted reward at a given state will remain the same. Crucially, unlike in \cite{hu2023_denseRL}, no pre-training or pre-solving of the problem is required to generate a dense reward, allowing the \gls{RL} to learn the optimal policy free of pre-determined, user-defined structure or reference trajectory.

The hyperparameters used for the \gls{RL} training are given in Table~\ref{table:RLparameters}.
\begin{table}[h!]
\centering
\caption{{\color{black}\gls{RL} parameters.}}
\label{table:RLparameters}
\adjustbox{max width=\linewidth}{
\begin{tabular}{lcccc} 
    \toprule
    Parameter & \multicolumn{4}{c}{All Scenarios} \\ [0.5ex] 
    \midrule
    Learning rate $\alpha_0$ & \multicolumn{4}{c}{$3$e$-4$}\\
    Clipping rate $\epsilon_0$ & \multicolumn{4}{c}{0.2}\\
    Initial stochasticity, $\sigma_0$ & \multicolumn{4}{c}{0.1}\\
    Batch-size (episodes) & \multicolumn{4}{c}{25 }\\
    Epochs per update & \multicolumn{4}{c}{10 }\\
    \midrule
    Parameter & GTOC 11 & Earth-Mars & Psyche & 67P \\ [0.5ex] 
    \midrule
    Time-step, $\delta t$ & 30 days &  8.71975 days & 0.025 rev & 0.025 rev\\
    Average steps/episode & 60 & 40 & 75 & 50 \\
    \bottomrule
\end{tabular}
}
\end{table}

\section{\label{sec:Results}Results}

In this section we present the results of training the \glspl{gcnet} with \gls{BC} and \gls{RL} for the four transfer scenarios. First, we compare the computational load required by both approaches. Next, we present the nominal state performance for both the interplanetary rendezvous and small-body landing scenarios. Next we subject use the trained \glspl{gcnet} with a \gls{ZOH} on the control output, and explore their robustness to disturbances in \gls{IC}, \gls{OD} and \gls{EX}.

\subsection{\label{sec:ComputationalComparison}Training comparison}

\begin{figure*}[!t]
  \centering
  \includegraphics[width=0.48\textwidth]{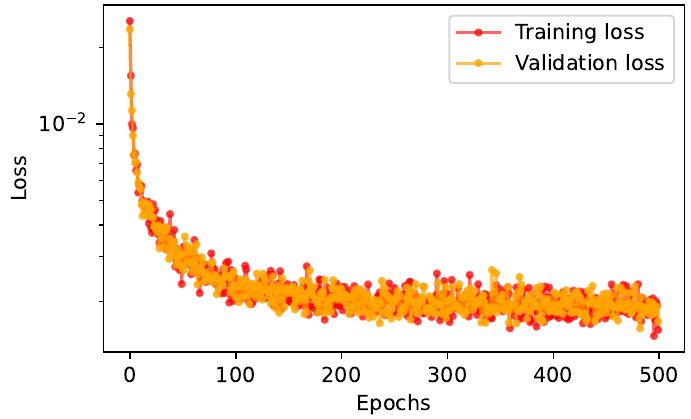}
  \hfill
  \includegraphics[width=0.48\textwidth]{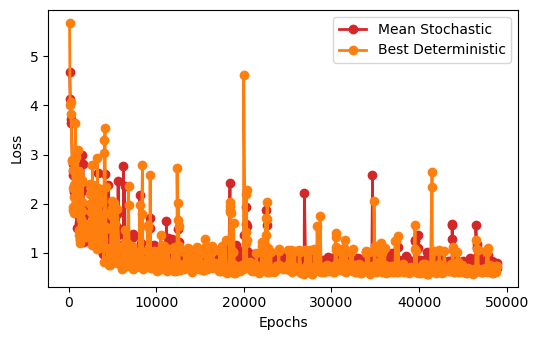}
  \caption{Training and validation loss of \gls{BC} (left) and \gls{RL} (right) \glspl{gcnet} for case A (\textit{GTOC 11}).}
  \label{fig:losses_figs}
\end{figure*}

The training and validation loss during training are depicted in Fig.\ref{fig:losses_figs} for the case A (\textit{GTOC 11}). Table~\ref{table:ComputationalTimeComparison} shows a comparison on the number of training samples required to obtain the best found \gls{gcnet} for each scenario. For the \gls{BC}, the size of the training dataset remains similar across the scenarios and are generated in a matter of seconds once the \gls{TPBVP} is solved using the \gls{bgoe} technique\cite{izzo2021,IzzoOriger2023_GCNETTime}. The \gls{gcnet} can be trained in around 1-3 hours. The \gls{RL} varies more across the scenarios considered, and takes approximately 1-24 hours to train. This will depend on the chosen batch size, architecture and parallelisation used. Case A (\textit{GTOC 11}) requires many more samples to learn, perhaps because it is the most challenging problem to solve given the control authority present and the size of the target event (see Table~\ref{table:taxonomy}). It is also computationally the slowest despite the quick integration time needed for a single episode given the constant thrust and lack of a lambert arc grid search. The discontinuous thrust profile and lambert arcs make the fuel-optimal transfers more time consuming to train. The \gls{BC}, on the other hand, is less susceptible to the variations in problem difficultly. However, the \gls{BC} requires a database of optimal/expert samples, whereas the \gls{RL} doesn't and instead generates its own, mostly sub-optimal, samples.

\begin{table}[h!]
\centering
\caption{Comparing the sample efficiency for \gls{BC} and \gls{RL}. Table shows the number of samples seen during training.}
\label{table:ComputationalTimeComparison}
\adjustbox{max width=\linewidth}{
\begin{tabular}{ccc cc} 
    \toprule
    Approach & GTOC 11 & Earth-Mars & Psyche & 67P \\
    & $\times10^6$ & $\times10^6$ & $\times10^6$ & $\times10^6$\\
    \midrule
    \gls{BC} & 40 & 35 & 35 & 24 \\
    \gls{RL} & 634 & 24 & 27 & 11 \\
    \bottomrule
\end{tabular}
}
\end{table}

\subsection{\label{sec:Nominal}Nominal Results}

To begin with, we compare the performance of the \glspl{gcnet} on the nominal initial conditions with continuous integration of the \glspl{gcnet} inside the taylor-adaptive integrator.

\subsubsection{\label{sec:InterplanetaryNom}Interplanetary Rendezvous}

\begin{figure*}[htb]
  \centering
  \includegraphics[width=0.48\textwidth]{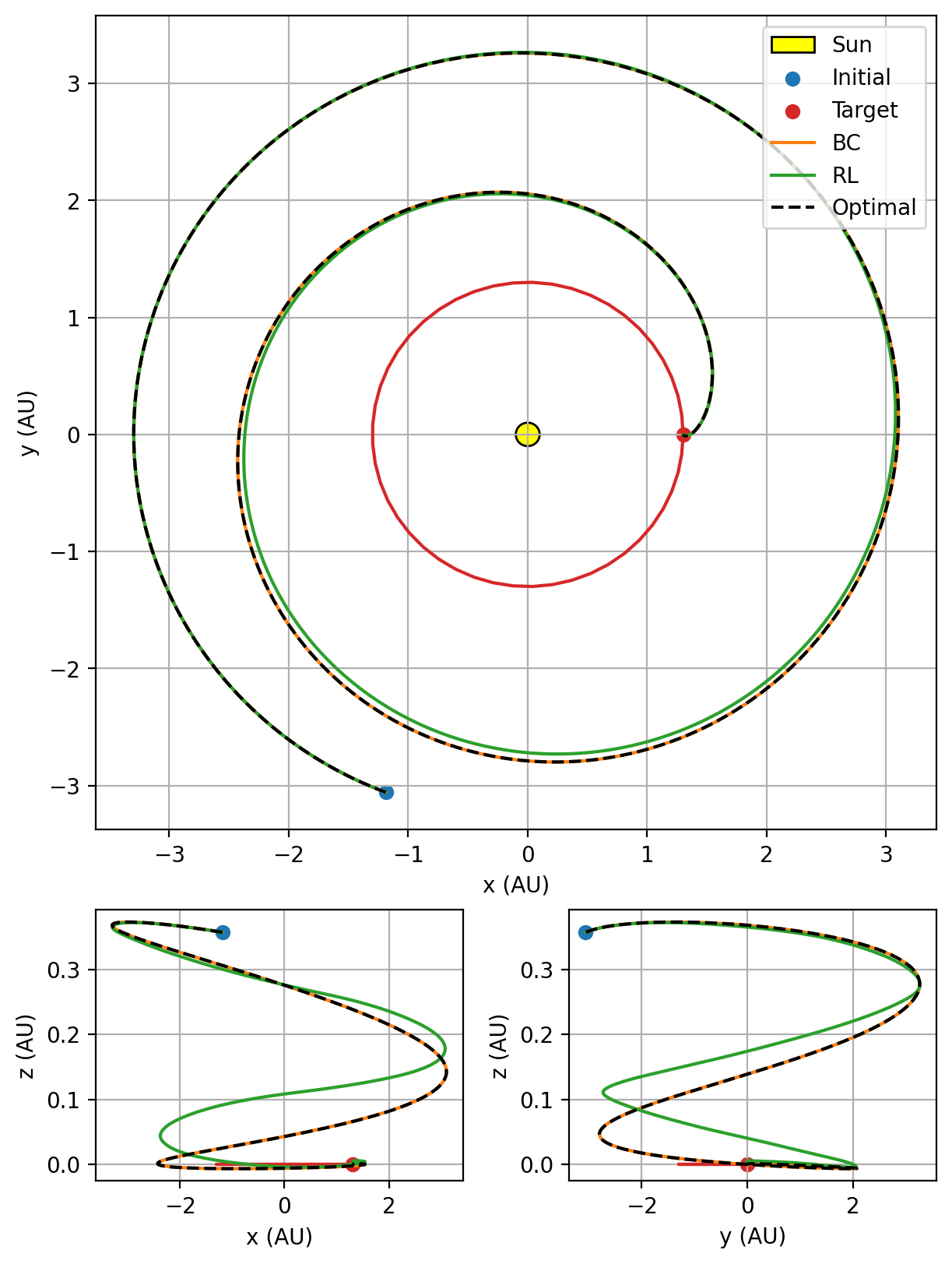}
  \hfill
  \includegraphics[width=0.48\textwidth]{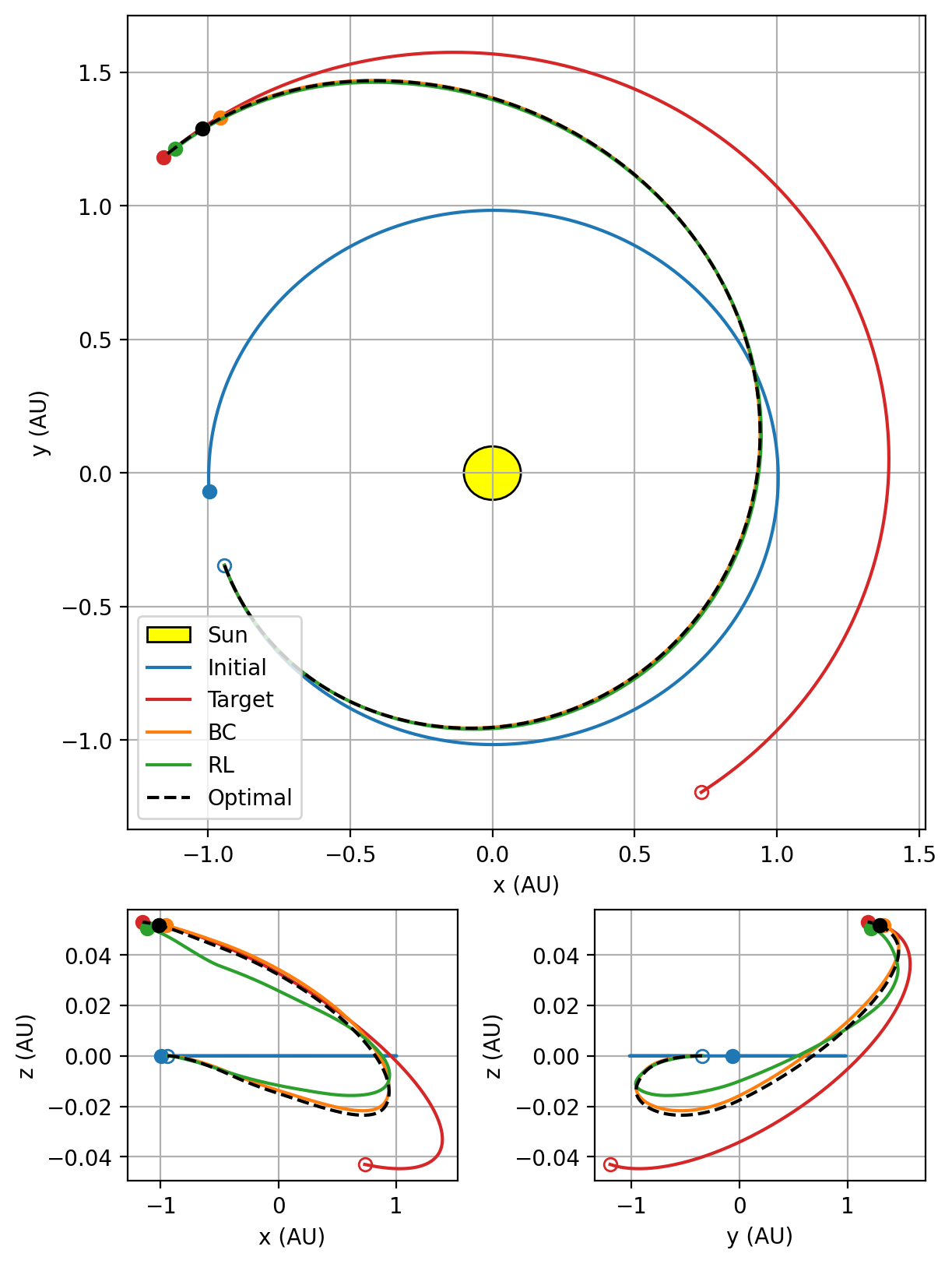}
  \caption{Nominal \gls{BC} and \gls{RL} \glspl{gcnet} for case A (\textit{GTOC 11}) (left) and B (\textit{Earth-Mars})  (right) in the rotating and inertial frames respectively, with the Optimal control solution shown for comparison.}
  \label{fig:Interplanetary_Determ_Results}
\end{figure*}

Table~\ref{table:Interplanetary_Determ_Results} compares the two \glspl{gcnet} to the solution obtained by solving the \gls{TPBVP} with an indirect method (labelled \textit{Optimal}). In both case A (\textit{GTOC 11}) and B, an event function at the \gls{SOI} is used. As such, the value of the \textit{optimal} solution at the event is also given. For the time-optimal case A (\textit{GTOC 11}), the \gls{BC}-\gls{gcnet} is only $0.30\%$ away from the optimal solution, losing out on 5 days over 4.5 years. It also enters the \gls{SOI} with a lower velocity residual to the target compared to the optimal solution. The \gls{RL}-\gls{gcnet} takes an extra $14.8$ days to reach the \gls{SOI}, corresponding to $0.91\%$ of the total time-of-flight. Figure~\ref{fig:Interplanetary_Determ_Results} (left) gives a visual comparison of the trajectories resulting from following the optimal, \gls{BC}-\gls{gcnet} and \gls{RL}-\gls{gcnet} control profiles. As expected, it is hard to distinguish between the optimal and \gls{BC}-\gls{gcnet} trajectories, indicating the \gls{BC} approach is accurately replicating the optimality principles. The \gls{RL}-\gls{gcnet} deviates slightly in the \textit{xy} plane, and a larger discrepancy is seen along the \textit{z}-axis. 

Whilst case A (\textit{GTOC 11}) is computed in an inertial reference frame, this is only possible given the circular target orbit. Case B (\textit{Earth-Mars}) represents an alternative scenario, where the target orbit is eccentric and therefore the transfer is computed in the inertial reference frame. Figure~\ref{fig:Interplanetary_Determ_Results} (right) compares the trajectories resulting from following the optimal, \gls{BC}-\gls{gcnet} and \gls{RL}-\gls{gcnet} control profiles. In this case, the differing arrival times correspond to different target locations. The \gls{BC}-\gls{gcnet} appears to struggle to replicate the nominal optimal control solution as well as in the time-optimal case. It arrives $5.72$ days earlier and saves $14.5$ kg of the fuel. However, this comes at the expense of arriving with a significantly higher velocity residual, $1501$ m/s, compared to $962$ m/s of the optimal. The \gls{RL}-\gls{gcnet}, in comparison, arrives $9.88$ days later and uses $23.2$ kg more fuel. However, the advantage is it arrives inside the \gls{SOI} with a lower velocity residual $380$ m/s. This is an advantage of the Lambert arc grid search described in Section~\ref{sec:RL_LossMass}, where it is not only trying to minimise the $\Delta v$ to reach the \gls{SOI} but also account for the remaining $\Delta v$ required to reach the target. If this were not the case, the velocity residual would be closer to the largest permissible during training, $c_v=1000$m/s.

\begin{table}[h!]
\centering
\caption{Interplanetary Rendezvous Nominal Results}
\label{table:Interplanetary_Determ_Results}
\adjustbox{max width=\linewidth}{
\begin{tabular}{lll ccccc} 
    \toprule
    Case & Objective & Approach & Time-of-flight & Spacecraft Mass & \multicolumn{2}{c}{Optimality Residual} & Velocity Residual\\
    \cmidrule{6-7}
    & & & [-] & [$m_f$/$m_0$] & [-] & [$\%$] & [m/s] \\
    \midrule
    GTOC 11 & Time & Optimal & 4.6194 years & 1 & - & - & -\\
    \cmidrule{3-8}
    & & Optimal @ Event & 4.4809 years & 1 & - & - & 421.90 \\
    & & \gls{BC} \gls{gcnet} & 4.4946 years & 1 & 5.0 days & 0.30 & 386.04 \\ 
    & & \gls{RL} \gls{gcnet} & 4.5215 years & 1 & 14.8 days & 0.91 & 440.44 \\
    \midrule
    Earth-Mars & Mass & Optimal & 348.79 days & 0.6039 & - & - & -\\
    \cmidrule{3-8}
    & & Optimal @ Event & 335.02 days & 0.6343 & - & - & 962.29\\
    & & \gls{BC} \gls{gcnet} & 329.30 days & 0.6488 & -14.5 kg & -2.29 & 1501.55 \\ 
    & & \gls{RL} \gls{gcnet} & 344.90 days & 0.6110 & 23.2 kg & 3.66 & 379.57 \\ 
    \bottomrule
\end{tabular}
}
\end{table}

\subsubsection{\label{sec:SmallBodyNom}Small-body Landing}

\begin{figure*}[htb]
  \centering
  \includegraphics[width=\textwidth]{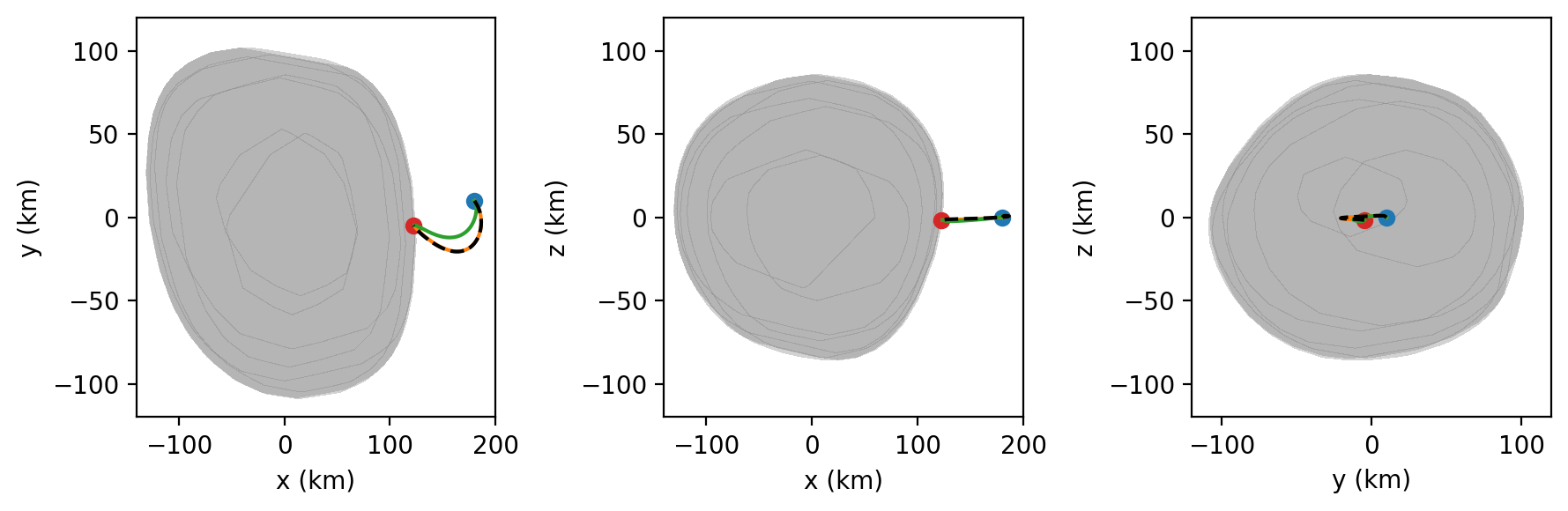}
  \hfill
  \includegraphics[width=\textwidth]{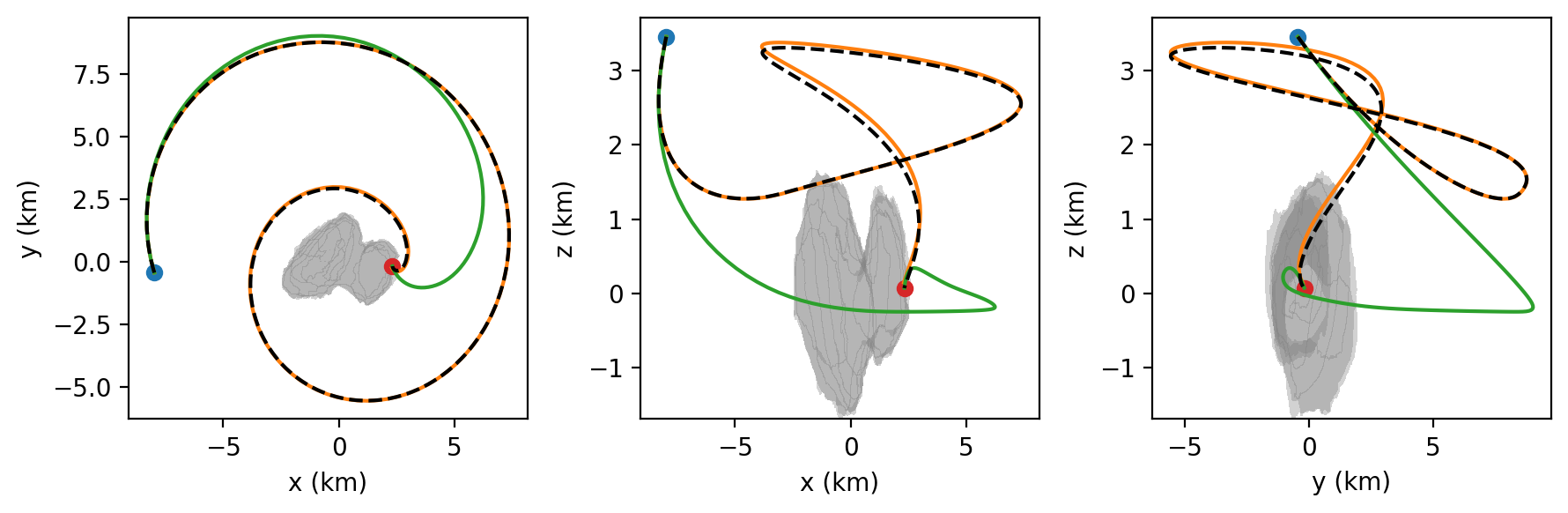}
  \caption{\gls{RL} \glspl{gcnet} for Psyche (top) and 67P (bottom) in the rotating frame.}
  \label{fig:Smallbody_Determ_Results}
\end{figure*}

Table~\ref{table:Smallbody_Determ_Results} compares the two \glspl{gcnet} to solutions obtained by solving the \gls{TPBVP} with an indirect method (labelled \textit{Optimal}) for the two fuel-optimal small-body landing scenarios. In this case, the surface of each small-body is represented by a \gls{NN}-event as described in~\ref{sec:Events}. Unlike for case B (\textit{Earth-Mars}) above, here the \gls{BC}-\gls{gcnet} replicates the nominal optimal solution well, only losing $0.12\%$ of fuel-optimality. In addition, the position and velocity residuals are even better than the optimal nominal solution. In contrast the \gls{RL}-\gls{gcnet} is less optimal, using $0.84\%$ more fuel, however with a smaller position and velocity residual. This is highly noticeable in Fig.~\ref{fig:Smallbody_Determ_Results} where the \gls{RL} trajectory takes a more direct line to the target compared to the nominal. We suggest this is because the \gls{RL} is in a local minima where arriving at the event earlier means less fuel consumption.

However, the story is noticeably different for the landing on 67P. This case proved very challenging to solve the optimal control problem, with the additional constraint of avoiding the surface before converging to the target. In the end a local optimum was found that could be used to generate the a suitable dataset for training the \gls{BC}-\gls{gcnet}. The optimal solution takes $15.739577$ hours to reach within $5$ m of the target, with a $0.023$m/s velocity residual. The \gls{BC}-\gls{gcnet} is less accurate with a state residual of  $102.13$m and $0.119$m/s when reaching the comet surface, but uses $0.016\%$ less fuel. The \gls{RL}-\gls{gcnet}, on the other hand, finds a very different trajectory, as seen in Fig.~\ref{fig:Smallbody_Determ_Results}. Instead of $1.5$ revolutions in the rotating frame, it uses $0.5$ revolutions. This corresponds to $11.557893$ hours of flight, and uses $0.035\%$ less fuel whilst also meeting the position constraint of $5$m and having a lower velocity residual that the indirect solution found. To confirm this, we also closed the final state reached by the \gls{RL}-\gls{gcnet} to the target with the indirect method and found the combined trajectory to be more optimal than the original indirect (local) optimal solution found. This indicates a local-minimum solution was found and used for the \gls{BC} training. A more rigorous search for the true optimal solution would, of course, improve upon the \gls{RL}-\gls{gcnet} solution and also lead to a better \gls{BC}-\gls{gcnet}. However, it would require additional work to incorporate the comet surface constraint whilst solving the \gls{TPBVP} and indicates the complexity of the search space. \gls{RL} avoids this by exploring the environment and shows potential as a means of generating initial guesses for optimal control solutions.

\begin{table}[h!]
\centering
\caption{Small-body Landing Nominal Results}
\label{table:Smallbody_Determ_Results}
\adjustbox{max width=\linewidth}{
\begin{tabular}{lll cccccc} 
    \toprule
    Case & Objective & Approach & Time-of-flight & Spacecraft Mass & \multicolumn{2}{c}{Optimality Residual} & Position Residual & Velocity Residual\\
    \cmidrule{6-7}
    & & & [hours] & [$m_f$/$m_0$] & [-] & [$\%$] & [m] & [m/s] \\
    \midrule
    Psyche & Time & Optimal & 0.822918  & 0.944029 & - & - & - & -\\
    \cmidrule{3-9}
    & & Optimal @ Event & 0.782900  & 0.960661 &  - & - & 1772.50 & 24.47 \\
    & & \gls{BC} \gls{gcnet} & 0.787306 & 0.958558 & 0.74 & 0.22 & 1430.80 & 22.74 \\
    & & \gls{RL} \gls{gcnet} & 0.602728 & 0.946440 & 5.03 & 1.48 & 1167.68 & 20.93 \\
    \midrule
    67P & Mass & Optimal (local) & 15.857123 & 0.997804 & - & - & -& -\\
    \cmidrule{3-9}
    & & Optimal (local) @ Event & 15.739577 & 0.997850 & - & - & 5.00 & 0.023 \\
    & & \gls{BC} \gls{gcnet} & 15.461748 & 0.998006 & -0.016 & -0.016 & 102.13 & 0.119 \\
    & & \gls{RL} \gls{gcnet} & 11.557893 & 0.998204 & -0.035 & -0.035 & 5.00 & 0.016 \\

    \bottomrule
\end{tabular}
}
\end{table}

\subsection{\label{sec:Stochastic}Stochastic Results}

So far the results presented all start from the nominal initial states indicated in Tables~\ref{table:TestCase} and~\ref{table:TestCase_Asteroid}, and integrate the \gls{gcnet} continuously in the right-hand side of the dynamical equations - see Eq.~\eqref{eq:IntegrateRHS}. However, the presence of uncertainties in state and thrust execution are a major concern for autonomous spacecraft operations. As such, we test the performance of the \glspl{gcnet} subject to uncertainties in \gls{IC}, \gls{OD}, and \gls{EX} using several 200-samples Monte Carlo simulations.
\noindent First, whilst integrating the \glspl{gcnet} continuously in the right-hand side of the dynamical equations, they are subject to:
\begin{itemize}
    \item \gls{IC}: uniform errors with a $\pm \Delta r_{OI}$ and $\pm \Delta v_{OI}$ in the position and velocity components about the nominal \glspl{IC}. An initial mass error of $\Delta m_{OI}$ is also added for the small-body landings. 
\end{itemize}
We implement a \gls{ZOH} on the control outputs from the \glspl{gcnet}, lasting $\delta t_{ZOH}$. Again, these are subject to:
\begin{itemize}
    \item A missed-thrust probability per time-step of $p_{ZOH}$, with a duration $\Delta p_{ZOH}$.
    \item \gls{OD}: uniform errors with a $\pm \Delta r_{OD}$ and $\pm \Delta v_{OD}$ in the position and velocity components are added every $\delta t_{OD}$ time-steps.
    \item \gls{EX}: uniform spherical errors of magnitude $\pm \Delta T \%$ lasting $\delta t_{EX}$ time-steps.
\end{itemize}
Table~\ref{table:StochasticParameters} shows the values used for the various error magnitudes across the different test cases.


\begin{table}[h!]
\centering
\caption{Stochastic Errors Introduced}
\label{table:StochasticParameters}
\adjustbox{max width=\linewidth}{
\begin{tabular}{ll cccc} 
    \toprule
    Name & Variable & GTOC 11 & Earth-Mars & Psyche & 67P \\ [0.5ex] 
    \midrule
    \gls{IC} & $\Delta r_{OI}$ & 500,000 km & 100,000 km & 165 m & 4.5 km\\
    & $\Delta v_{OI}$ & 250 m/s & 50 m/s & 8.5 m/s  & 0.5 m/s \\
    & $\Delta m_{OI}$ & - & $0.0\%$ & $10.0\%$  & $5.0\%$ \\
    \midrule
    \gls{ZOH} & $\delta t_{ZOH}$ & $1$ day & $1$ day & $15$ s & $1$ min \\
    & $p_{ZOH}$ & 1/365.25 & 1/365.25 & 1/15 & 1/90\\
    & $\Delta p_{ZOH}$ & 7 days & 7 days & 1 min & 5 min \\
    \midrule
    \gls{OD} & $\Delta r_{OD}$ & 50,000 km & 10,000 km & 25 m & 5 m \\
    & $\Delta v_{OD}$ & 25 m/s & 5 m/s & 1 m/s & 0.1 m/s \\
    & $\delta t_{OD}$ & 28 days & 28 days & 1 min & 5 min \\
    \midrule
    \gls{EX} & $\Delta T$  & 10 \% & 10 \% & 5 \% & 5 \%  \\
    & $\delta t_{EX}$ & 28 days & 28 days & 1 min & 5 min\\
    \bottomrule
\end{tabular}
}
\end{table}

\subsubsection{\label{sec:InterplanetaryStoch}Interplanetary Rendezvous}

\begin{figure*}[htb]
  \centering
  \includegraphics[width=0.48\textwidth]{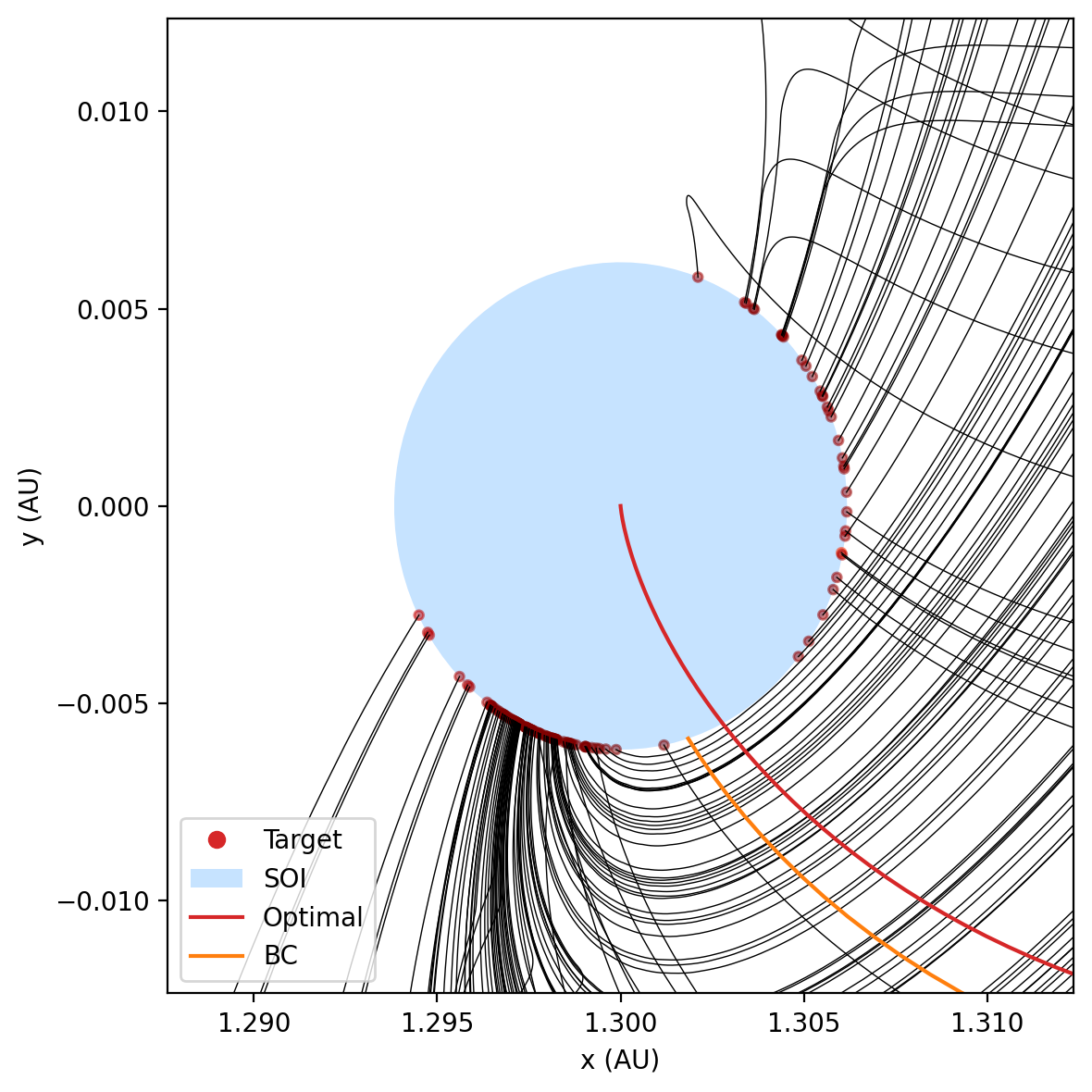}
  \hfill
  \includegraphics[width=0.48\textwidth]{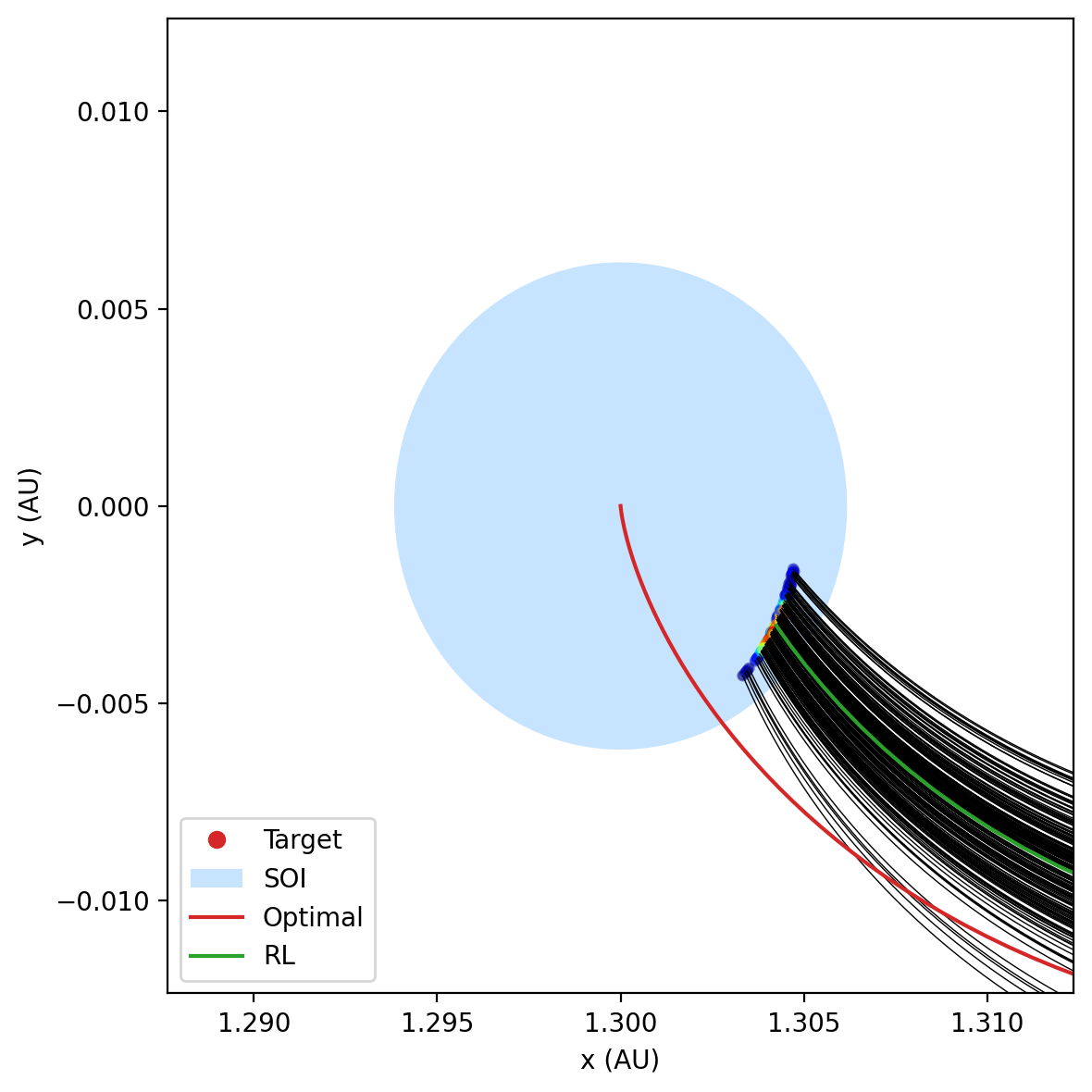}
  \caption{\glspl{gcnet} performance for case A (\textit{GTOC 11}) subject to stochastic \glspl{IC} for \gls{BC} (left) and \gls{RL} (right) in the rotating frame.}
  \label{fig:gcnet_comparison_caseA_dist}
\end{figure*}

Table~\ref{table:Interplanetary_Stoch_Results} shows the percentage of trajectories that converge to the target in position only, $r$, and full state, $x$, for each set of stochastic error realisations. The same stochastic seed is used to compare \gls{BC} and \gls{RL}. For case A (\textit{GTOC 11}), the \gls{BC}-\gls{gcnet} and  \gls{RL}-\gls{gcnet} are comparable with each other, handling each of the \gls{IC}, \gls{ZOH}, \gls{OD} and \gls{EX} errors well. Figure~\ref{fig:gcnet_comparison_caseA_dist} shows the bundle of trajectories subject to \gls{IC} errors at the \gls{SOI}. It's clear the \gls{RL} achieves a tighter bundle and is more robust to \gls{IC} errors. During training, the \gls{RL}-\gls{gcnet} is subject to stochastic \glspl{IC} and the objective is to ensure they all converge to the target. It will therefore trade optimality to achieve this higher level of robustness.

\begin{figure*}[htb]
  \centering
  \includegraphics[width=0.48\textwidth]{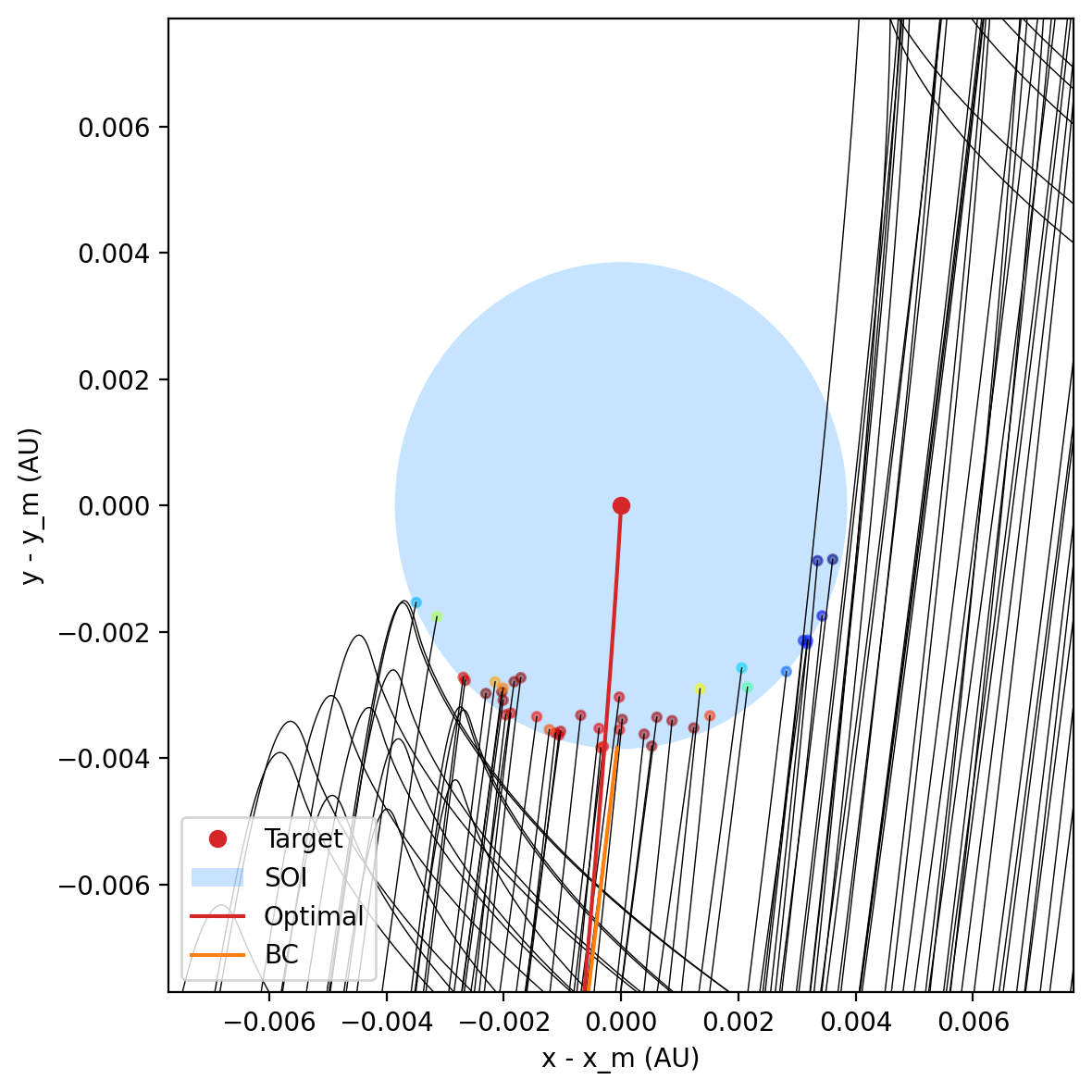}
  \hfill
  \includegraphics[width=0.48\textwidth]{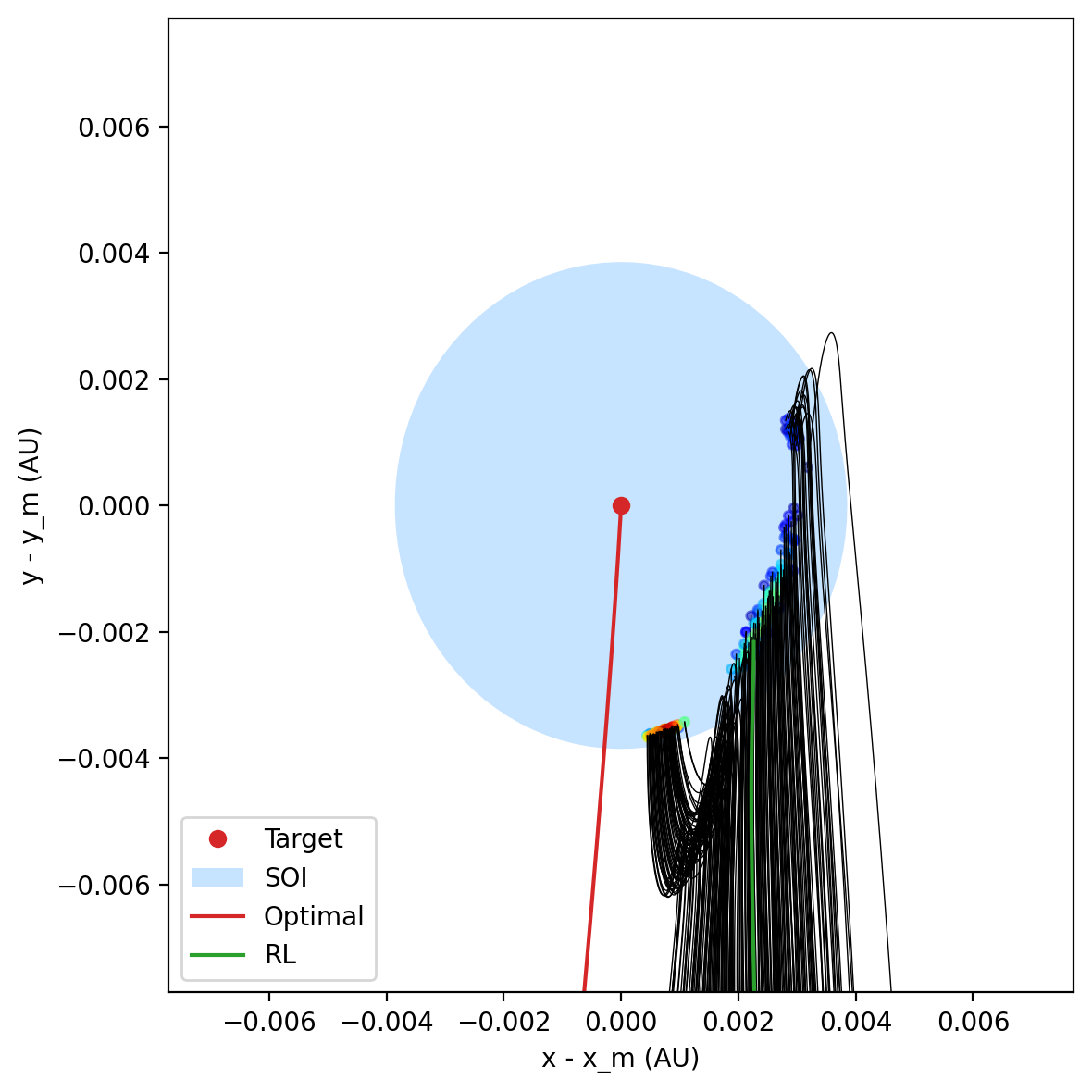}
  \caption{\glspl{gcnet} performance for case B (\textit{Earth-Mars}) subject to stochastic \glspl{IC} for \gls{BC} (left) and \gls{RL} (right) in the inertial frame.}
  \label{fig:gcnet_comparison_caseB_dist}
\end{figure*}

In case B (\textit{Earth-Mars}), the inertial reference frame and the nature of the moving target proves challenging for the \gls{BC} approach. The bundle of trajectories used in the \gls{bgoe} database has a fixed target location after the target time-of-flight. It is not aware of the targets location at other time steps. In contrast, the \gls{RL} is allowed to arrive at any time less than or equal to the target time-of-flight. Hence, it experiences different arrival locations at different epochs during training, and hence is more robust to stochastic errors that change the arrival time. This is clearly demonstrated in the distribution of arrival trajectories at Mars once the \glspl{IC} are subject to errors as seen in Fig.~\ref{fig:gcnet_comparison_caseB_dist}. In addition, the very nature of the \gls{PPO} training, where stochastic actions and \glspl{IC} force exploration of the environment, prepare the \gls{gcnet} for unseen stochastic errors such as \gls{ZOH} and \gls{OD}. Future work can look to improve \gls{BC} performance by adding trajectories to the database of expert examples.

\begin{table}[h!]
\centering
\caption{Interplanetary Rendezvous Stochastic Evaluation}
\label{table:Interplanetary_Stoch_Results}
\adjustbox{max width=\linewidth}{
\begin{tabular}{lll cc cc cc cc} 
    \toprule
    Case & Objective & Approach & \multicolumn{2}{c}{\gls{IC}} & \multicolumn{2}{c}{\gls{ZOH}} & \multicolumn{2}{c}{\gls{OD}} & \multicolumn{2}{c}{\gls{EX}}\\
    \cmidrule{3-5}\cmidrule{6-7}\cmidrule{8-9}\cmidrule{10-11}
    & & Converged & $r$ & $x$ & $r$ & $x$ & $r$ & $x$ & $r$ & $x$ \\
    & & & [$\%$] & [$\%$] & [$\%$] & [$\%$] & [$\%$] & [$\%$] & [$\%$] & [$\%$] \\
    \midrule
    GTOC 11 & Time & \gls{BC} \gls{gcnet} & 100 & 100 & 99.5 & 98.5 & 100 & 98 & 88 & 84.5\\
    & & \gls{RL} \gls{gcnet} & 100 & 100 & 93 & 93 & 100 & 100 & 100 & 98.5 \\
    \midrule
    Earth-Mars & Mass & \gls{BC} \gls{gcnet} & 20.5 & 4 & 69.5 & 2.5 & 100 & 7.5 & 0.5 & 0.5 \\
    & & \gls{RL} \gls{gcnet} & 100 & 100 & 100 & 93.5 & 100 & 100 & 94 & 87 \\
    \bottomrule
\end{tabular}
}
\end{table}

\subsubsection{\label{sec:SmallBodyStochastic}Small-body Landing}

\begin{figure*}[htb]
  \centering
  \includegraphics[width=0.48\textwidth]{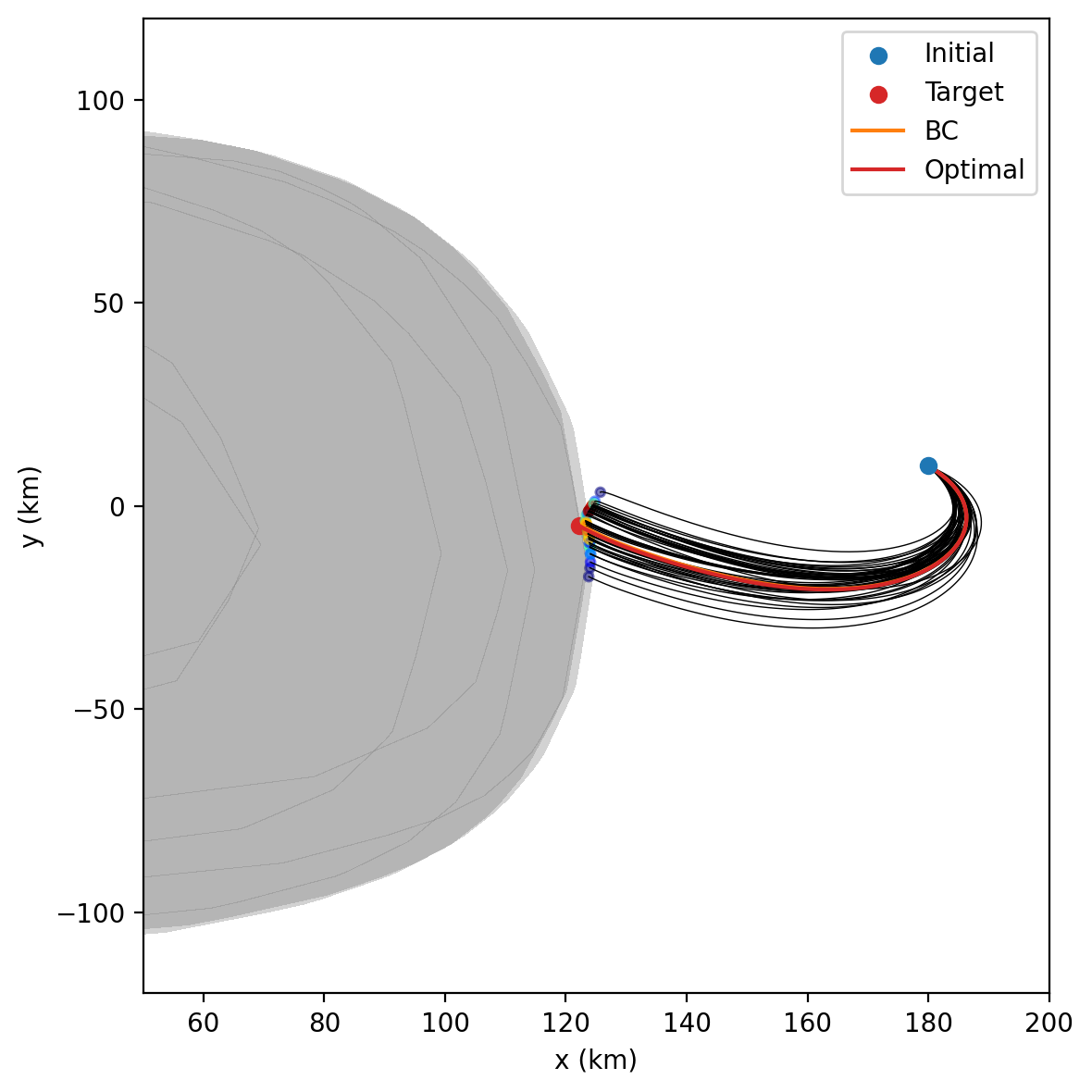}
  \hfill
  \includegraphics[width=0.48\textwidth]{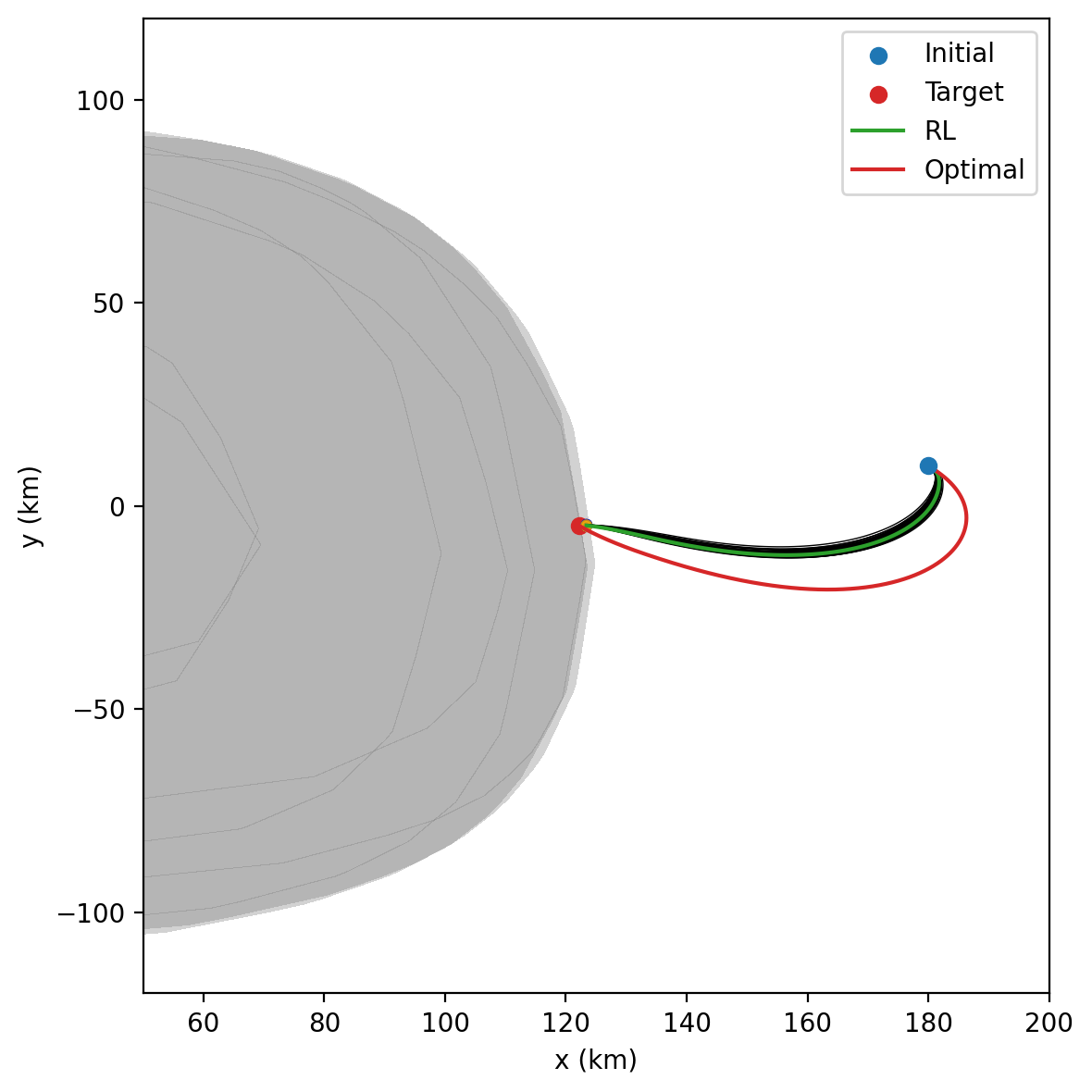}
  \caption{\glspl{gcnet} performance for Psyche subject to stochastic \glspl{IC} for \gls{BC} (left) and \gls{RL} (right) in the rotating frame.}
  \label{fig:gcnet_comparison_Psyche_dist}
\end{figure*}

\begin{figure*}[htb]
  \centering
  \includegraphics[width=0.48\textwidth]{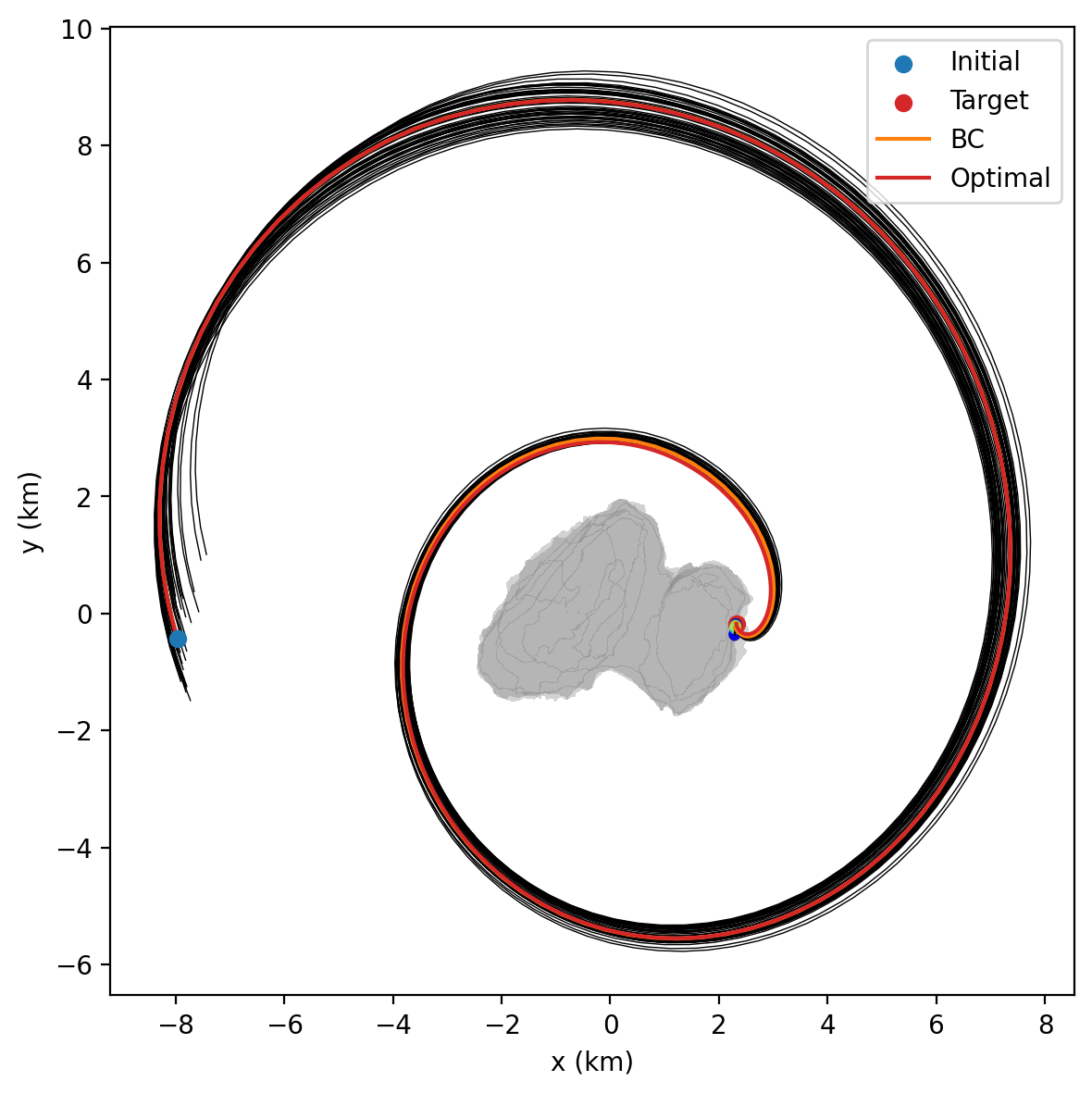}
  \hfill
  \includegraphics[width=0.48\textwidth]{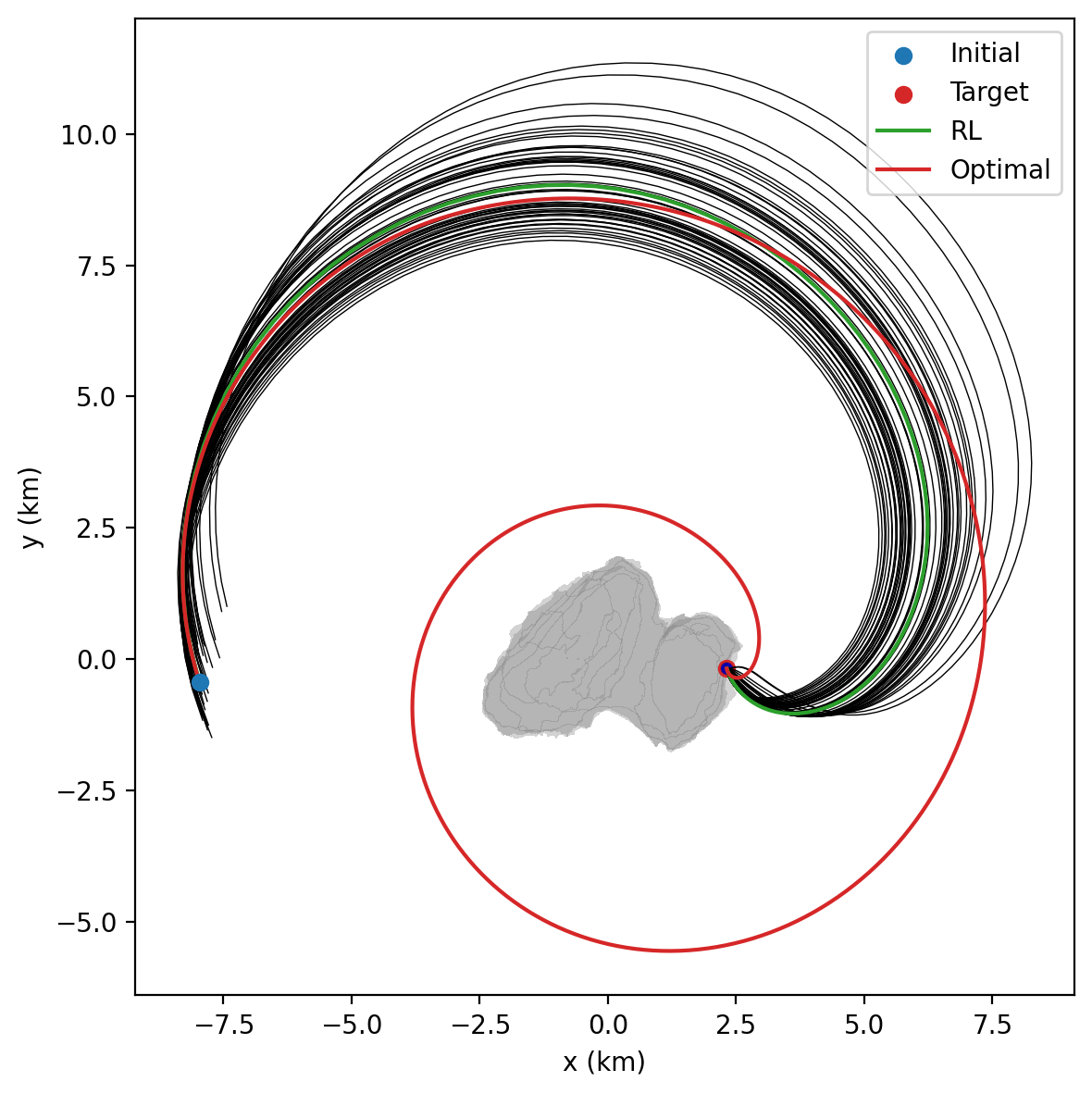}
  \caption{\glspl{gcnet} performance for 67P subject to stochastic \glspl{IC} for \gls{BC} (left) and \gls{RL} (right) in the rotating frame.}
  \label{fig:gcnet_comparison_67P_dist}
\end{figure*}

A similar story emerges in the small-body landing scenarios. Although both transfers are now in rotating reference frames, the \gls{RL} out-performs the \gls{BC} in the majority of cases. Figures~\ref{fig:gcnet_comparison_Psyche_dist} and~\ref{fig:gcnet_comparison_67P_dist} show the distribution of trajectories obtained from \glspl{gcnet} subject to \gls{IC} errors for both Psyche and 67P. The results are summarised in Table~\ref{table:Smallbody_Stoch_Results}.

The \gls{RL} consistently outperforms the \gls{BC} in the presence of the same realisation of stochastic errors. For Psyche, the \gls{BC} struggles with the \gls{IC} and \gls{EX} errors in particular. The velocity error in the \gls{IC} simulations is quite large, and could cause the larger distribution seen in Fig.~\ref{fig:gcnet_comparison_Psyche_dist}, particularly along the $y$-axis, which is the direction in which the surface is moving in this orientation of the landing site. Again the \gls{RL} results in a much tighter bound on the trajectory bundle. For 67P, however, the \gls{ZOH} and \gls{OD} errors cause difficulty for the \gls{BC}-\gls{gcnet}. Although not plotted, \gls{OD} errors cause particularly large deviations, well outside the bundle of trajectories used in the training database - see Fig.~\ref{fig:bgoe_case67P}, and result in $0\%$ convergence to the target state.

\begin{table}[h!]
\centering
\caption{Small-body Stochastic Evaluation}
\label{table:Smallbody_Stoch_Results}
\adjustbox{max width=\linewidth}{
\begin{tabular}{lll cc cc cc cc} 
    \toprule
    Case & Objective & Approach & \multicolumn{2}{c}{\gls{IC}} & \multicolumn{2}{c}{\gls{ZOH}} & \multicolumn{2}{c}{\gls{OD}} & \multicolumn{2}{c}{\gls{EX}}\\
    \cmidrule{3-5}\cmidrule{6-7}\cmidrule{8-9}\cmidrule{10-11}
    & & Converged & $r$ & $x$ & $r$ & $x$ & $r$ & $x$ & $r$ & $x$ \\
    & & & [$\%$] & [$\%$] & [$\%$] & [$\%$] & [$\%$] & [$\%$] & [$\%$] & [$\%$] \\
    \midrule
    Psyche & Time & \gls{BC} \gls{gcnet} & 22 & 22 & 100 & 64.5 & 96.5 & 82 & 0 & 0 \\
    & & \gls{RL} \gls{gcnet} & 100 & 100 & 100 & 92.5 & 100 & 100 & 100 & 100 \\
    \midrule
    67P & Mass & \gls{BC} \gls{gcnet} & 100 & 100 & 19 & 19 & 0 & 0 & 100 & 100 \\
    & & \gls{RL} \gls{gcnet} & 100 & 100& 96 & 96& 74.5 & 74.5 & 100 & 100 \\
    \bottomrule
\end{tabular}
}
\end{table}


\section{\label{sec:Conclusion}Conclusion}

\begin{table}[tb]
\centering
\caption{Summary of \gls{gcnet} training methodologies}\label{table:summary}
\adjustbox{max width=\linewidth}{
\begin{tabular}{lll}
\toprule
Approach & Pro/Con & Comment \\
\midrule
\gls{BC} & Pros & Computationally fast to train \\
& & Replicates expert behaviour well \\
& & More optimal on nominal conditions \\
\cmidrule{2-3}
& Cons & Requires generating a dataset of expert behaviour \\
& & Only as good as the expert dataset \\
& & Difficult to embed robustness \\
\midrule
\gls{RL} & Pros & No requirement to pre-solve the problem \\
& & More robust to errors (specifically out-of-distribution)\\
& & Can indicate non-intuitive solutions \\
\cmidrule{2-3}
& Cons & Computationally slower to train \\
& & Less likely to be optimal on nominal conditions\\
& & Requires reward function tuning\\
\bottomrule
\end{tabular}
}
\end{table}

\glsresetall

\Glspl{gcnet} are an increasingly viable alternative to existing on-board guidance and control approaches for spacecraft trajectory design. This paper presents a comprehensive comparison the two main training philosophies: \gls{BC} and \gls{RL}, in the context of spacecraft trajectory design and guidance problems. A similarly broad selection of problems with relatively unchanged \gls{gcnet} setups has not previously been considered in the literature, allowing a more general reflection of \gls{BC} and \gls{RL} for spacecraft transfers. We confirm what is already hypothesised in the literature, that \gls{BC} can provide more optimal solutions around the nominal initial conditions. However, \gls{RL} offers better out-of-distribution performance whilst preserving a degree of optimality, in essence rather than \lq\lq optimising a given objective function better, they intrinsically define a better objective\rq\rq\cite{song2023reaching}. Table~\ref{table:summary} summarises the advantages and disadvantages of \gls{RL} and \glspl{BC} for spacecraft \glspl{gcnet}.

Future work should not neglect either \gls{BC} or \gls{RL} approaches for spacecraft \gls{gcnet} design. The importance of both optimality and robustness means both have a role to play. We envisage techniques for improving the robustness of \gls{BC}-\glspl{gcnet} through Neural-ODE corrections~\cite{chen2018neural}, whilst \gls{RL}-\glspl{gcnet} could use \gls{BC} as a warm-start for optimality before adding robustness by exploring stochastic and unknown environments.


\bibliography{lib.bib}

\end{document}